\documentclass[prb,showpacs,preprintnumbers,amsmath,amssymb,twocolumn,superscriptaddress]{revtex4}
\usepackage[pdftex]{graphicx}
\usepackage{dcolumn}
\usepackage{bm}
\usepackage{color}

\begin{document}
\title{Spin fluctuations and superconductivity 
in layered $f$-electron superlattices
}

\author{Yasuhiro Tada}
\email[]{tada@issp.u-tokyo.ac.jp}
\affiliation{Institute for Solid State Physics, The University of
  Tokyo, Kashiwa 277-8581, Japan}
\author{Robert Peters}
\affiliation{Computational Condensed Matter Physics Laboratory, RIKEN, Wako, Saitama 351-0198, Japan}

\newcommand{\vecc}[1]{\mbox{\boldmath $#1$}}

\begin{abstract}
We investigate magnetic and superconducting properties of layered $f$-electron superlattices
within the fluctuation exchange approximation (FLEX). 
We show that spin fluctuations, which are characterized by the
maximum value of the spin susceptibility in the 3-dimensional (3D) Brillouin zone, 
are strongly suppressed in $f$-electron superlattices.
However, effective 
2D spin fluctuations can be increased 
due to the spatial confinement of the $f$-electrons.
Therefore, the tendency towards $d_{x^2-y^2}$-wave superconductivity,
mediated by these spin fluctuations,
can be strongly increased in $f$-electron-superlattices.
This is
in sharp contrast to superlattices composed of 
conventional $s$-wave superconductors,
where superconductivity is generally suppressed.
\end{abstract}


\maketitle

\section{introduction}
Recent experimental realizations of layered superlattices,
CeIn$_3$/LaIn$_3$ and CeCoIn$_5$/YbConIn$_5$, have opened new possibilities
in the field of $f$-electron systems.~\cite{pap:Shishido2010,pap:Mizukami2012,
pap:Goh2012,pap:Shimozawa2014}
Due to a non-trivial interplay of strong correlations and tunable dimensionality,
novel phenomena have been observed in these $f$-electron
superlattices 
which have not been seen so far in 
 existing magnetic/superconducting
superlattices composed of weakly or non-interacting systems.
~\cite{book:Shinjo2002,book:ChangGiessen1985,pap:Jin1989,
pap:Barnerjee1984,
pap:Kanoda1986,pap:Grunberg1986,pap:Parkin1990,
pap:Parkin1991,pap:Yafet1987,pap:Bruno1995}

For example,
it has been found that magnetic properties of CeIn$_3(n)$/LaIn$_3(4)$ superlattices~\cite{pap:Shishido2010}
depend on the thickness of the CeIn$_3$-layer within the unit cell of the superlattice.
In bulk CeIn$_3$, the coherence temperature is $T_{\rm coh}\sim 50$(K) and
the N\'eel temperature is $T_N\simeq 10$(K) with an ordering vector
$\vecc{Q}=(\pi,\pi,\pi)$.~\cite{pap:Lawrence1980,pap:Knebel2001,
pap:Pfleiderer2009}
Remarkably,
it has been reported that the N\'eel temperature of the superlattice is suppressed 
when the width of the CeIn$_3$-layers, $n$, is reduced, 
and eventually approaches zero
for $n=2$.
At the same time, a linear temperature dependence in the in-plane resistivity is observed 
for $n=2$, suggesting that the dimensionality of the antiferromagnetic (AF)
spin fluctuations
is reduced from 3-dimensions in bulk CeIn$_3$ to 2-dimensions in the superlattice.
Such an anomalous behavior in the resistivity has never been observed in
 previous studies of magnetic superlattices, and would be characteristic for 
$f$-electron superlattices.

Furthermore, superconductivity in
CeCoIn$_5$/YbCoIn$_5$ superlattices 
has been investigated.~\cite{pap:Mizukami2012,
pap:Goh2012,pap:Shimozawa2014}
In  bulk CeCoIn$_5$, the coherence temperature is 
$T_{\rm coh}\sim 50$(K) and
the superconducting transition temperature is $T_c\sim 2.3$(K).\cite{pap:Pfleiderer2009,pap:Petrovic2001}
CeCoIn$_5$ is located near an AF quantum critical point 
and AF spin fluctuations are expected to be important for the normal state as well as for
superconductivity. In the bulk system, the AF spin fluctuations are especially 
strong around $\vecc{Q}=(\pi,\pi,\pi)$ due to
the nesting of the Fermi surface, and they can be characterized as 
3D-like.~\cite{pap:Stock2008,pap:Kawasaki2003,pap:Shishido2002}
It is generally considered that 
the superconductivity exhibits $d_{x^2-y^2}$-wave symmetry and is mediated by these AF spin fluctuations.
Experiments on CeCoIn$_5$/YbCoIn$_5$ superlattices have demonstrated that
superconductivity exists
even for thin CeCoIn$_5$-layers and that
the superconducting transition temperature, $T_c$,
is suppressed as the width of the CeCoIn$_5$-layers
in the unit cell is reduced. However, it must be noted that at the same time effects of disorder, which are estimated from 
the residual resistivity, are 
increased in thin CeCoIn$_5$-layers.
It is thus unclear, how the superconductivity behaves in  ``clean" $f$-electron superlattices.

Motivated by these experiments, 
there have been several theoretical studies. 
The effects of a possible Rashba-like 
spin-orbit coupling due
to local inversion symmetry breaking near the interfaces of the 
Ce-layers and the spacer layers~\cite{pap:Maruyama2012,
pap:Yoshida2013,pap:Maruyama2013,pap:Yoshida2014} have been investigated. When the Rashba-like 
interaction is sufficiently large,
the Pauli depairing effect is greatly 
suppressed and novel superconducting states might be stabilized when a magnetic field is applied.
In another theoretical study, the experimental data was analyzed based on
the Berezinskii-Kosterlitz-Thouless transition
by regarding the superlattice as a junction composed of a normal metal
and a superconductor.~\cite{pap:She2012}
If this junction picture is applicable to the $f$-electron superlattice, 
then superconductivity in the YbCoIn$_5$-layer would be strongly suppressed,
because of a large mismatch between the Fermi velocities of the CeCoIn$_5$-layer and 
the YbCoIn$_5$-layer, 
leading to a 2-dimensional superconductivity in the CeCoIn$_5$-layers.

In these previous studies neither electron correlations nor the
superlattice structure are explicitly considered.
However, these are two key ingredients in 
$f$-electron superlattices and
distinguish them from all the existing non-interacting superlattices 
and the bulk $f$-electron compounds.
To understand $f$-electron superlattices, it is necessary to clarify the impact of electron correlations and the superlattice 
structure, and also their possible interplay.
In two previous studies, the present authors already discussed the Kondo effect and quasi-particles properties
\cite{pap:Tada2013,pap:Peters2013} using the dynamical mean field theory (DMFT), which captures local strong 
correlations, but neglects non-local fluctuations.
This time, we analyze magnetic and superconducting properties of $f$-electron superlattices 
using the fluctuation exchange approximation (FLEX) in order to describe
spatially extended spin fluctuations.~\cite{pap:Yanase2003}
We use a periodic Anderson model (PAM)
which is defined on a superlattice.
This model can be considered as a minimal model to describe $f$-electron superlattices, because it takes into account both
the electron correlations and the superlattice structures.

This paper is organized as follows:
In Sec. \ref{sec:model}, we introduce our model, the FLEX approximation for the spin
fluctuations, and the
Eliashberg equations for the superconductivity.
Spin fluctuations are discussed in Sec. \ref{sec:spin fluctuations},
and the superconducting instability is examined in 
Sec. \ref{sec:superconductivity}.
Finally, in Sec. \ref{sec:summary} we shortly summarize this paper. 

\section{model}
\label{sec:model}
\begin{figure*}
\includegraphics[width=0.3\linewidth]{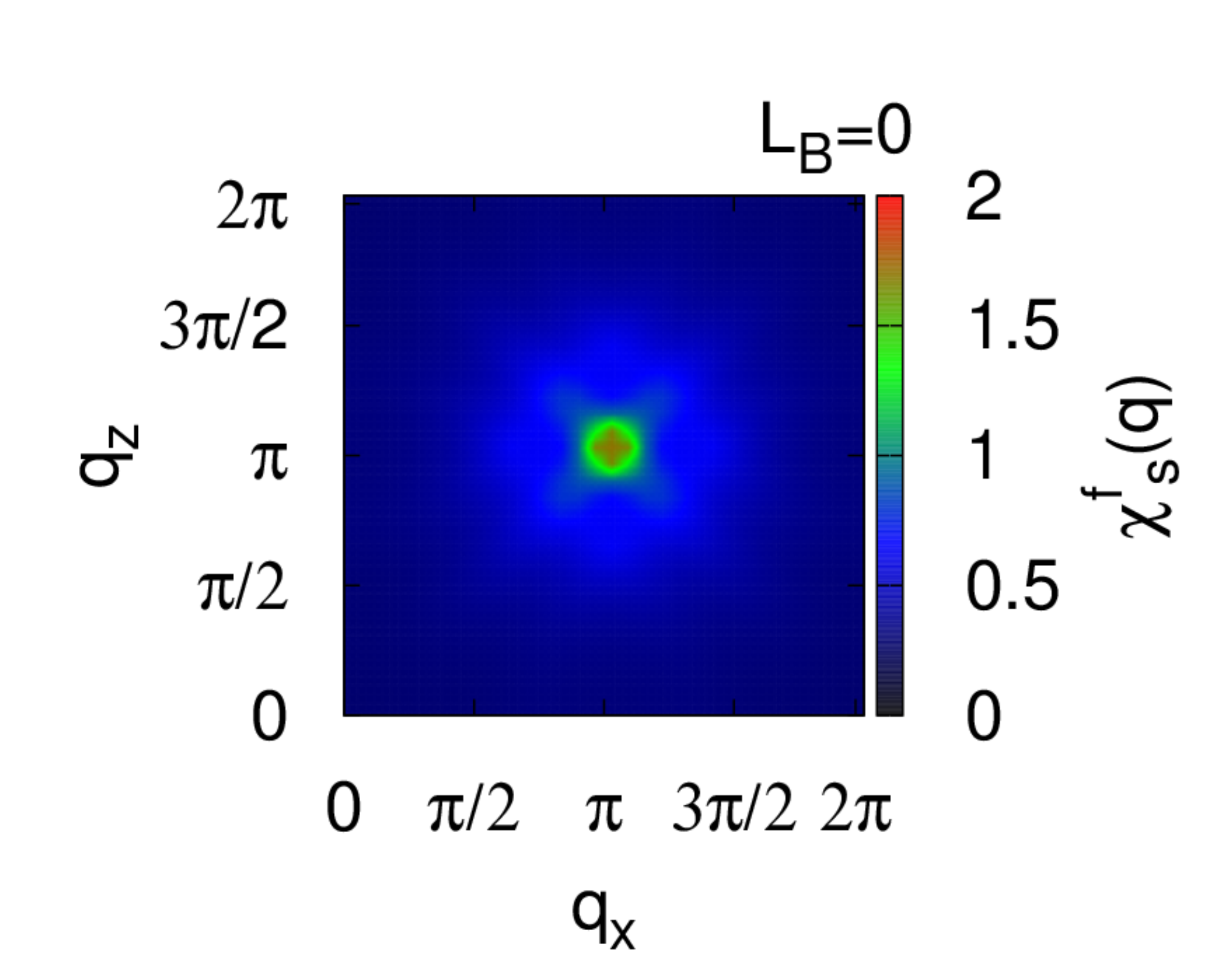}
\includegraphics[width=0.3\linewidth]{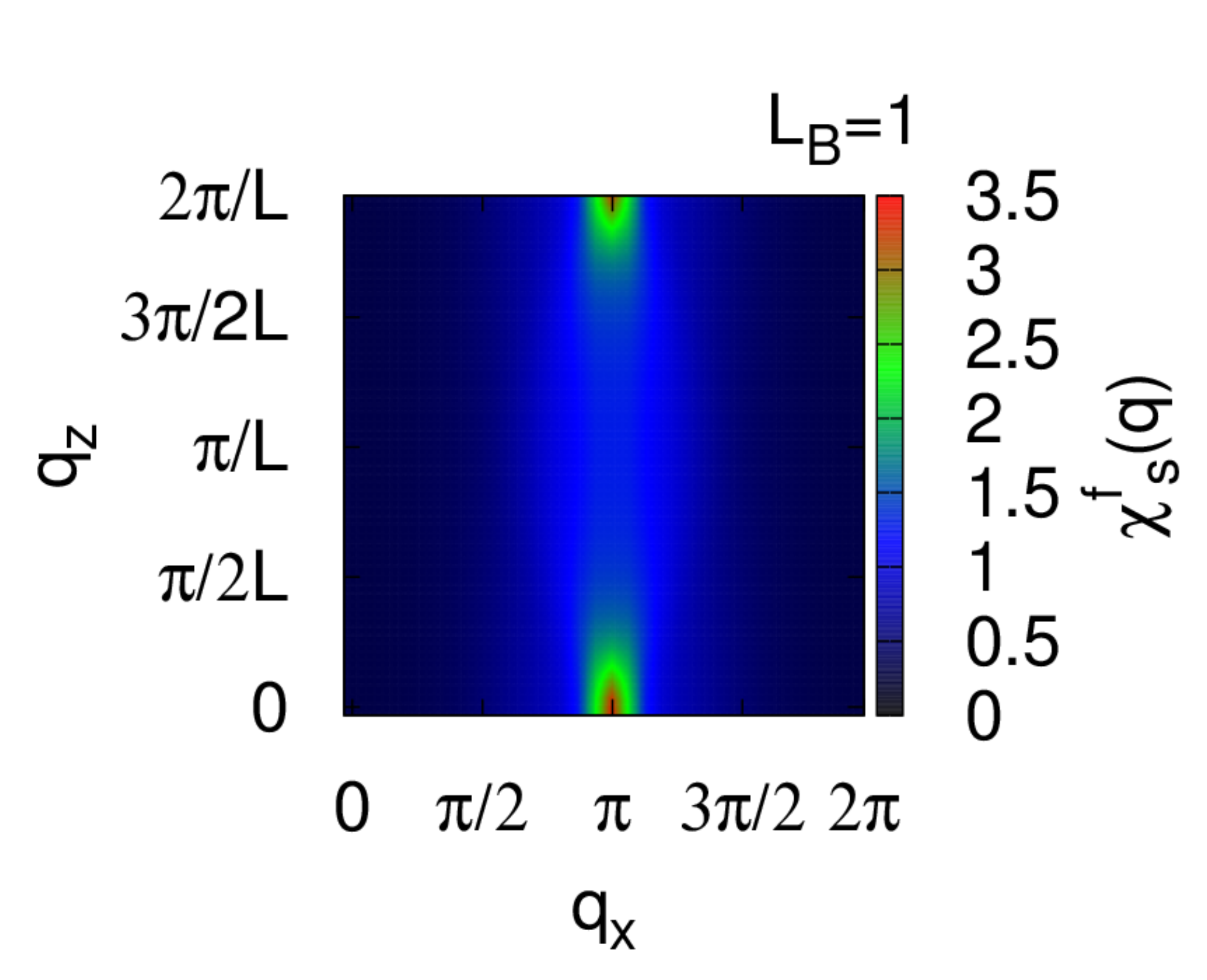}
\includegraphics[width=0.3\linewidth]{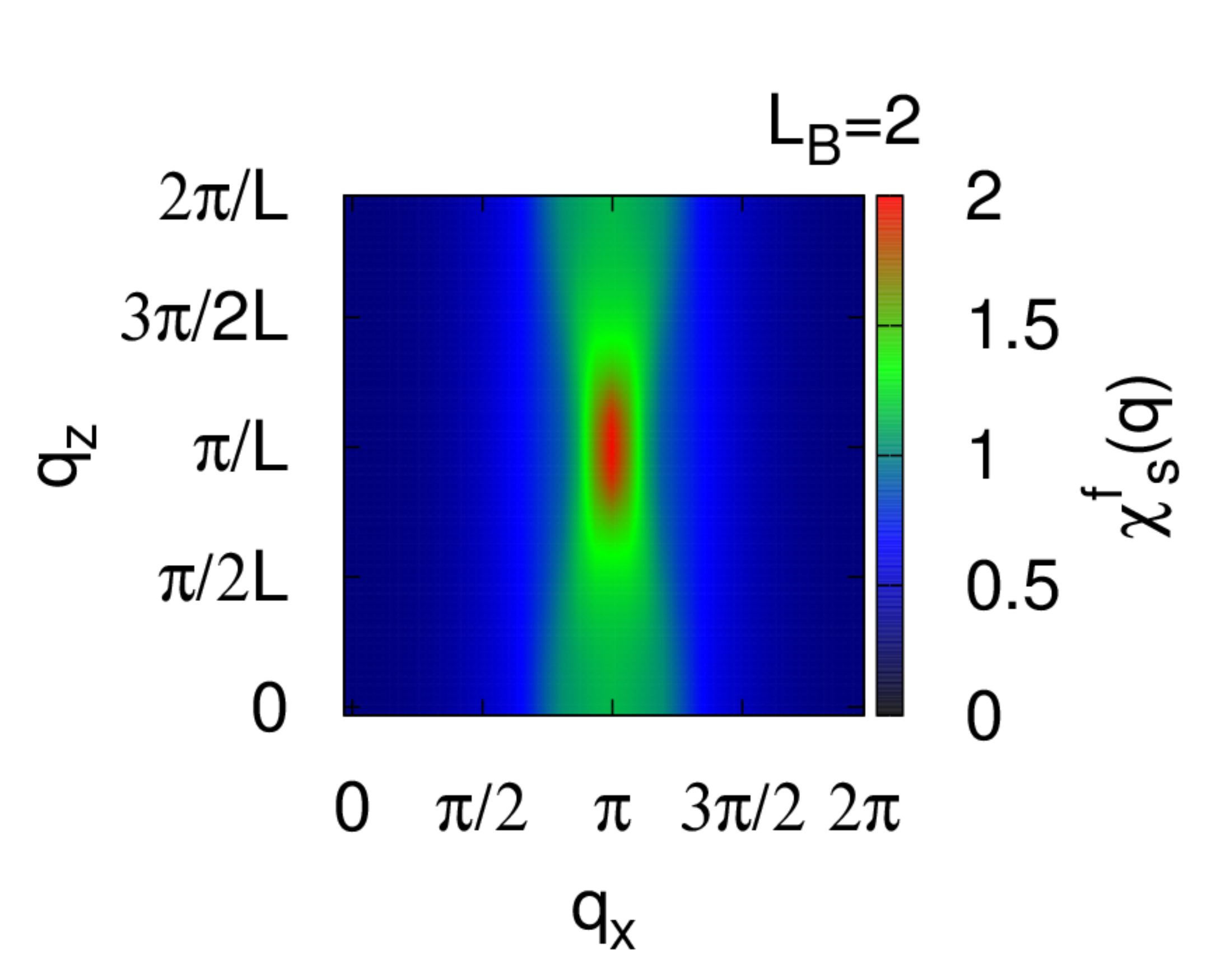}\\
\includegraphics[width=0.3\linewidth]{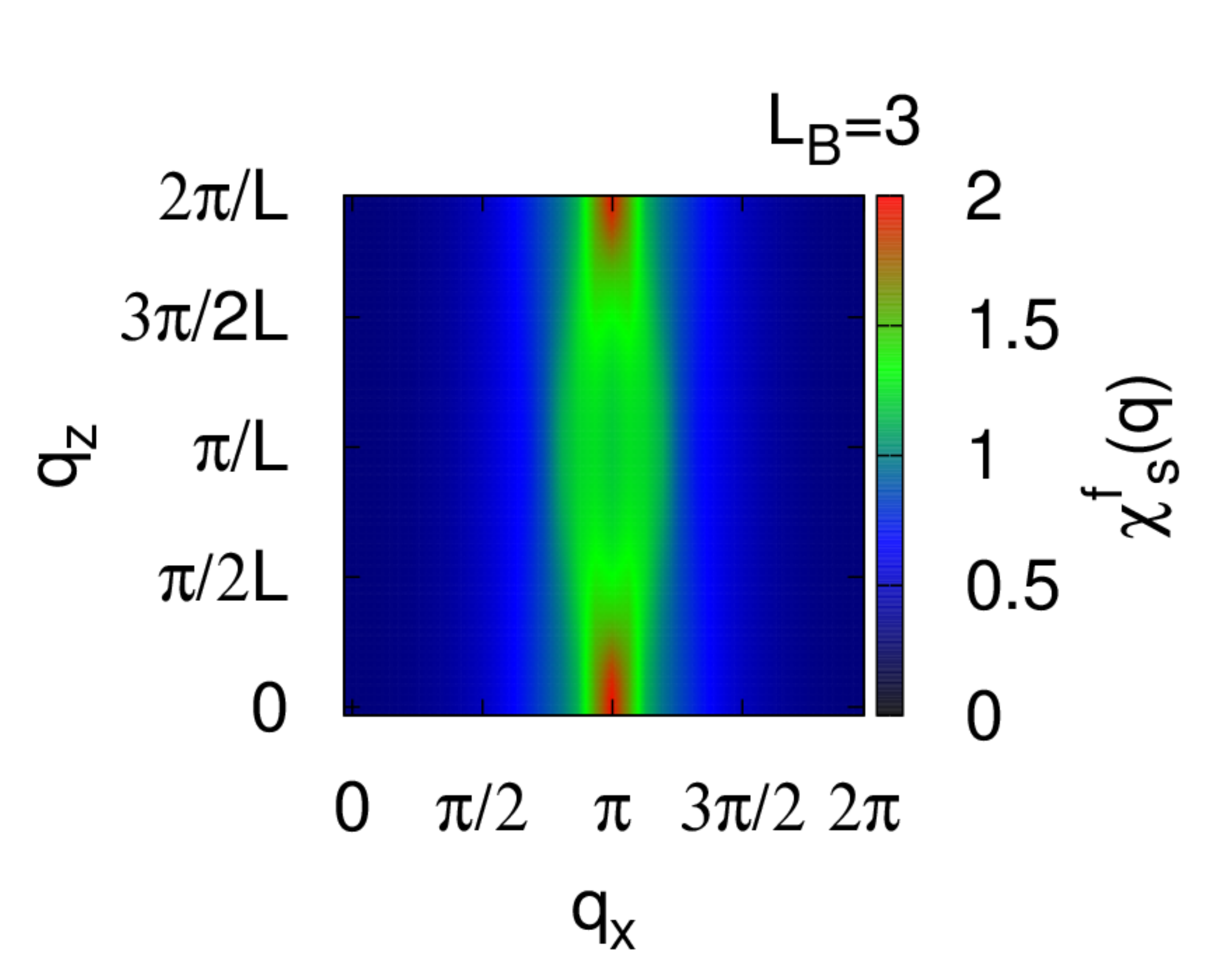}
\includegraphics[width=0.3\linewidth]{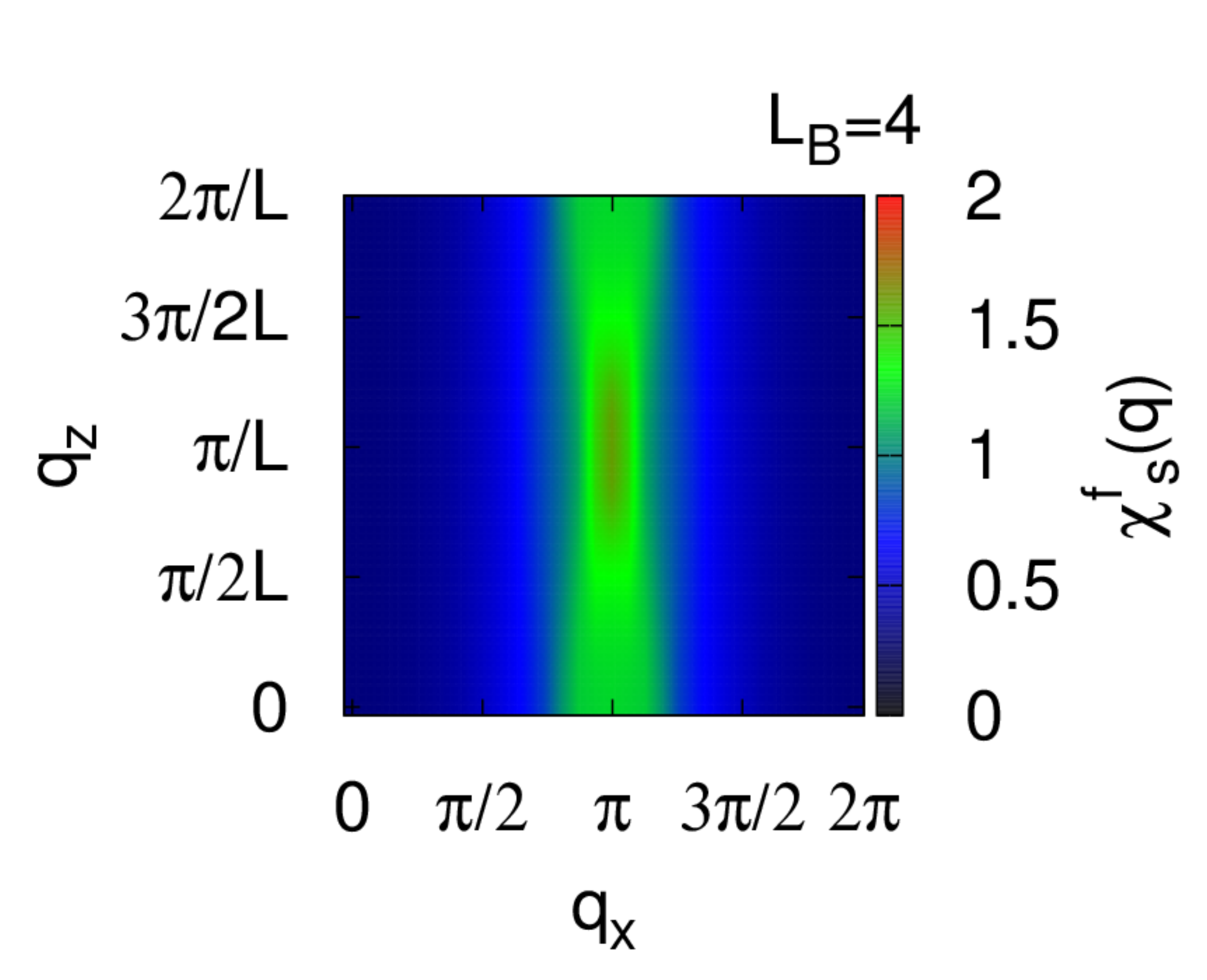}
\includegraphics[width=0.3\linewidth]{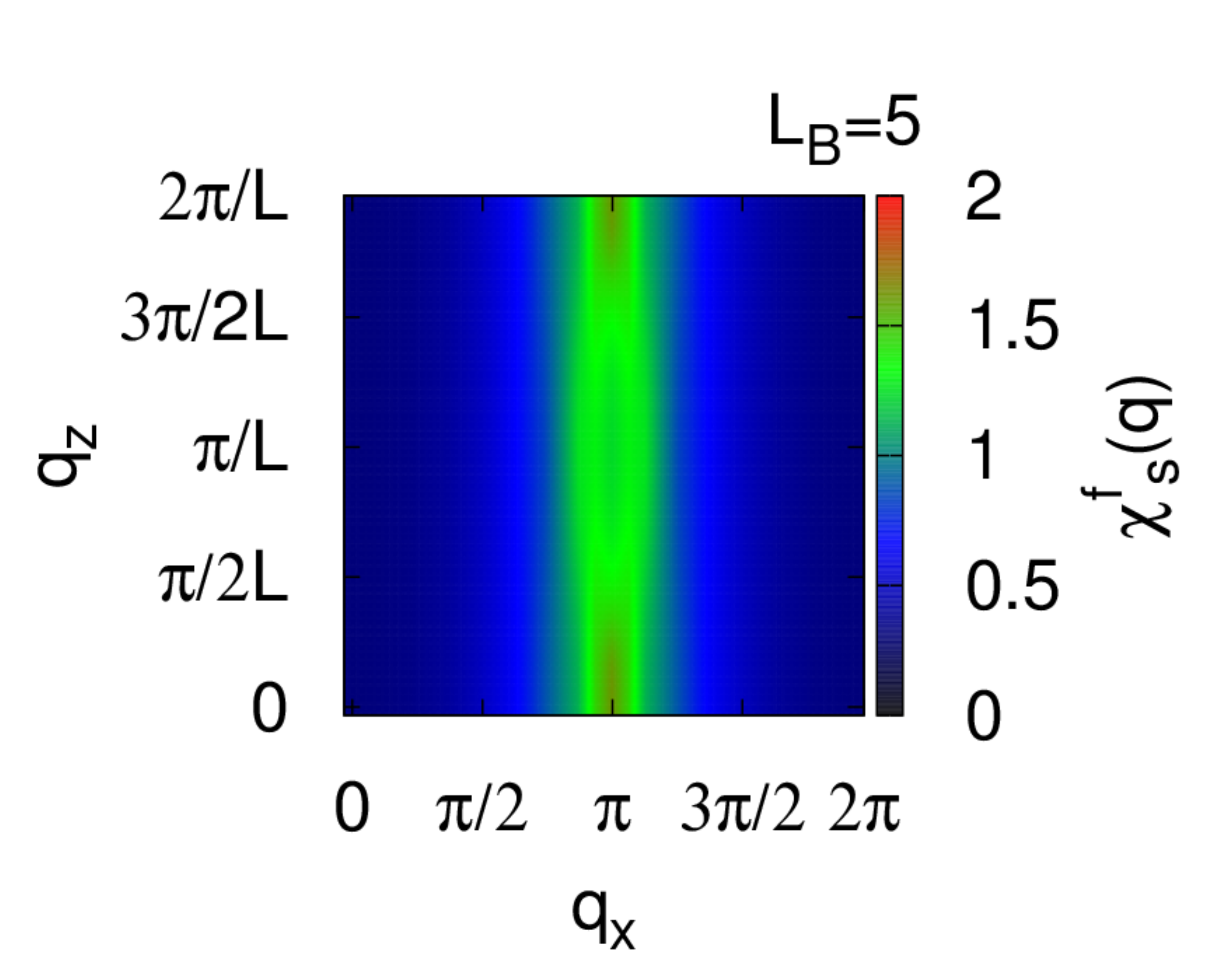}
\caption{
Spin susceptibility $\chi^f_s(i\omega_n=0,\vecc{q})$
at $T=0.1$
in the $xz$-plane at $q_y=\pi$ for $t^c_2=t^c_1$ for different $L_B$.
}
\label{fig:chisT01}
\end{figure*} 

Because the  $f$-electrons in the Yb-sites form a closed shell,
the YbCoIn$_5$-layers in CeCoIn$_5$/YbCoIn$_5$ superlattices
can be treated as normal uncorrelated metals.
Indeed, the resistivity in bulk YbCoIn$_5$ shows a monotonic 
temperature dependence without any signature of the Kondo effect.
Similarly, in the CeIn$_3$/LaIn$_3$ superlattices, 
only the CeIn$_3$-layers provide 
$f$-electrons at the Fermi energy.
Therefore, both superlattices can be considered as heterostructure composed of layers including $f$-electrons and layers without.
In order to understand these $f$-electron superlattices,
we introduce a PAM which consists of two kinds of 
layers, henceforth called
``A-layers'' and ``B-layers''.
The A-layers include conduction electrons ($c$-electrons) and $f$-electrons, which corresponds 
to CeIn$_3$- or CeCoIn$_5$-layers,
while within the B-layers only $c$-electrons exist, which corresponds to the 
LaIn$_3$- and YbCoIn$_5$-layers.
Our Hamiltonian thus reads
\begin{align}
H&=-\sum_{rr^{\prime}\sigma}t_{rr^{\prime}}^cc^{\dagger}_{r\sigma}c_{r^{\prime}\sigma}
-\sum_{rr^{\prime}\in {\rm A}, \sigma}t^f_{rr^{\prime}}
f^{\dagger}_{r\sigma}f_{r^{\prime}\sigma}\notag \\
&-\mu\sum_{r\sigma}c^{\dagger}_{r\sigma}c_{r\sigma}
-(\mu-\epsilon^f)\sum_{r\in {\rm A}, \sigma}
f^{\dagger}_{r\sigma}f_{r\sigma}\notag \\
&+V\sum_{r\in {\rm A},\sigma}[c^{\dagger}_{r\sigma}f_{r\sigma}
+f^{\dagger}_{r\sigma}c_{r\sigma}]
+U\sum_{r\in {\rm A}}n_{r\uparrow}^fn_{r\downarrow}^f,
\end{align}
where $c_{r\sigma}$ and $f_{r\sigma}$ are annihilation operators
for the conduction electrons and the $f$-electrons, respectively.
$r=(r_{\parallel},z)=(x,y,z)$ is a site index which is composed of an in-plane index , $r_{\parallel}$,
and a layer index $z$.
$\sigma$ corresponds to the spin index.
Each layer forms a square lattice, and hopping is only allowed between
nearest neighbor sites for simplicity;
$t^{a}_{r_{\parallel}zr_{\parallel}^{\prime}z}
=t^a_1$ for $r_{\parallel}\neq r_{\parallel}^{\prime}$
and $t^a_{r_{\parallel}zr_{\parallel}z^{\prime}}=t^a_2$ for
$z\neq z^{\prime}$ where $a=c,f$.
The number of A-layers and B-layers within the unit cell of the superlattice are given by
$L_A$ and $L_B$.
In the present study, we fix $L_A=1$ for which effects of the 
spatial confinement of the $f$-electrons are expected to be 
particularly strong.
Because of $t_2^f=0$ and the absence of a direct hopping between the A-layers, which are
separated by the B-layers,  
the $f$-electrons can move along 
the $z$-direction only through the B-layers.
The model parameters are chosen as $(t^c_1,V,\varepsilon_f)
=(5.0,2.0,0)$ and the total filling is fixed to
$n=n^c+n^f=0.95$, which are a reasonable set of parameters 
and similar to the ones used in
the previous DMFT study.~\cite{pap:Tada2013}
The interaction strength is fixed at a moderate value, $U=3.0$,
for which a clear divergence in the spin susceptibility of the 3D system ($L_B=0$)
is visible at low temperature.
The $z$-axis hopping $t^c_2$ characterizes an anisotropy of the system for $L_B=0$.

In order to analyze momentum-resolved properties, we perform a Fourier transform,
\begin{align}
c_{jz\sigma}&=\sum_{k_{\parallel}k_zl}
U^{c}_{j\tilde{z}_1\tilde{z}_2,k_{\parallel}k_zl}
c_{k_{\parallel}k_zl\sigma},\\
f_{jz\sigma}&=\sum_{k_{\parallel}k_z}
U^{f}_{j\tilde{z}_1\tilde{z}_2,k_{\parallel}k_z}
f_{k_{\parallel}k_z\sigma},
\end{align}
where the unitary matrices $U^c$ and $U^f$ are defined as,
\begin{align}
U^c_{j\tilde{z}_1\tilde{z}_2,k_{\parallel}k_zl}
&=\frac{e^{ik_{\parallel}r_{\parallel}}}{\sqrt{N_{\parallel}}}
\frac{e^{ik_zz+iq^c_l\tilde{z}_2}}{\sqrt{N_z}},\\
U^f_{j\tilde{z}_1\tilde{z}_2,k_{\parallel}k_z}
&=\frac{e^{ik_{\parallel}r_{\parallel}}}{\sqrt{N_{\parallel}}}
\frac{e^{ik_zz}}
{\sqrt{N_z/L}},
\end{align}
$\vecc{k}_{\parallel}=(k_x,k_y)$. 
The layer index $z$ is parametrized as
$z=L\tilde{z}_1+\tilde{z}_2$ with $0\leq \tilde{z}_2<L$ for $U^c$
and $\tilde{z}_2=1$ for $U^f$, and
$0\leq l< L$ for $U^c$,
$q^c_l=2\pi l/L$.
The momentum along the $z$-axis is defined within the reduced
Brillouin zone (RBZ), $0\leq k_z<2\pi/L$.
$N_{\parallel}$ is the total number of sites within a layer and 
$N_z$ is the total number of layers.
Thus, the total number of sites is given by $N=N_{\parallel}N_z$.

In FLEX, we focus on spatially extended spin fluctuations.
~\cite{pap:Yanase2003}
The selfenergy and susceptibilities in the normal state are given by,
\begin{align}
\Sigma^f(k)&=\frac{T}{N}\sum_{q}V^f(q)G^{f}(k-q),\\
V^f(q)&=\frac{1}{2}U^2\chi^f_c(q)+\frac{3}{2}U^2\chi^f_s(q)-U^2\chi^f_0(q)\\
\chi^f_0(q)&=-\frac{T}{N}\sum_q G^f(k+q)G^f(k),\\
\chi^f_c(q)&=\frac{\chi^f_0(q)}{1+U\chi^f_0(q)},\\
\chi^f_s(q)&=\frac{\chi^f_0(q)}{1-U\chi^f_0(q)}, \label{eq_sus}
\end{align}
where $k=(i\omega_n,\vecc{k})$
with $\vecc{k}=(\vecc{k}_{\parallel},k_z)$.
$G^f(k)$ is the $f$-electron Green's function in the $f_{k\sigma}$-basis.
Note that the $f$-electron contributions to the
total spin susceptibility are dominant, especially near magnetic criticality,
and that they are strongly enhanced by the interaction $U$.
On the other hand, 
contributions from the $c$-electrons are not enhanced by an interaction term
in the present model.

The superconducting instability is investigated within the linearized
Eliashberg equation for the singlet gap function 
$\Delta^f(k)$,
\begin{align}
\Delta^f(k)&=-\frac{T}{N}\sum_{k'} V^f_s(k-k')|G^{f}(k')|^2\Delta^f(k'),
\label{eq:Eliashberg}\\
V^f_s(q)&=U-\frac{1}{2}U^2\chi^f_c(q)+\frac{3}{2}U^2\chi^f_s(q).
\end{align}
It is noted that, $c$-electrons can only become superconducting via the $f$-electrons by the proximity effect,
because of the absence of $c$-electrons interactions.
The proximity effect is well taken into account in our calculations
because the $f$-electron Green's
function $G^f$ fully includes the hybridization processes between
the $f$-electrons and the $c$-electrons through $V$.

\section{spin fluctuations}
\label{sec:spin fluctuations}
\begin{figure}
\begin{center}
\includegraphics[width=0.8\hsize,height=0.4\hsize]{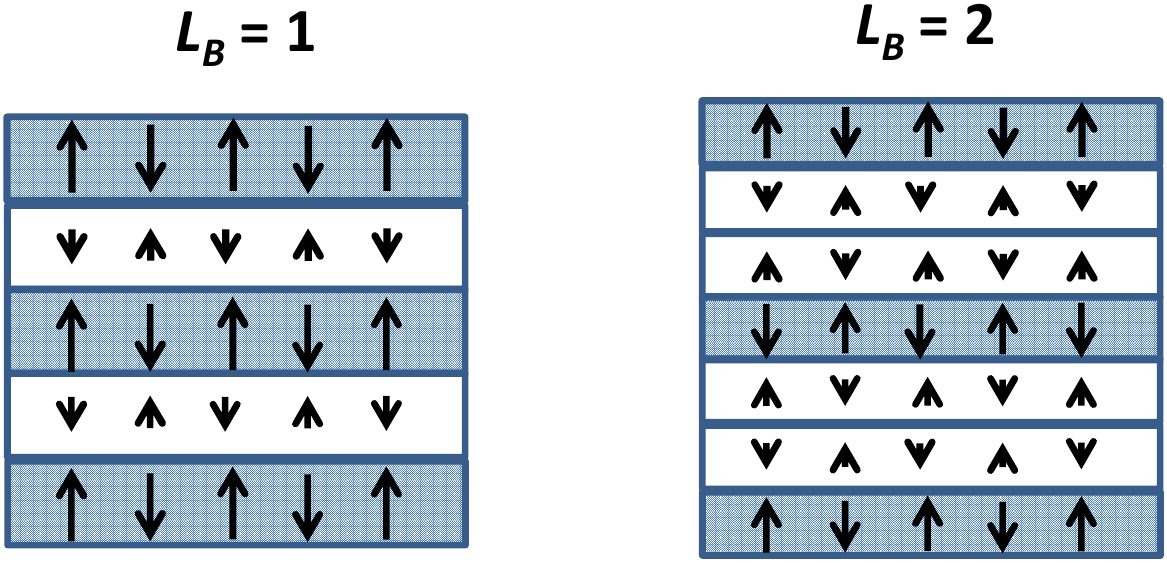}
\caption{
Schematic picture of the spin configurations for $L_B=1$ (left panel)
and $L_B=2$ (right panel).
Shaded layers and white layers are A-layers and B-layers, respectively.
}
\label{fig:config}
\end{center}
\end{figure}

In this section, we discuss the spin fluctuations as calculated by the FLEX.
First, we consider an isotropic parameter set, $t^c_2=t^c_1$, where
anisotropy between the $xy$- and $z$- directions
can only originate from the superlattice structure when $L_B\geq1$.
Figure \ref{fig:chisT01} shows the $q$-dependence of the magnetic susceptibility, $\chi^f_s
(i\omega_n=0,\vecc{q})$, 
at relatively high temperature, $T=0.1$.
The positions of the maximum values of $\chi^f_s$ oscillate depending on the number of spacer layers $L_B$.
When $L_B$ is even, the maximum values are located at $\vecc{Q}=(\pi,\pi,\pi/L)$;
when $L_B$ is odd they are at $\vecc{Q}=(\pi,\pi,0)$. 
These momenta correspond to the spin configurations shown
in Fig.~\ref{fig:config}.
When $L_B$ is odd (even), the magnetic coupling between different A-layers, which is mediated by the $c$-electrons, is ferromagnetic (antiferromagnetic).
Due to the proximity to the A-layers,
small moments are induced into the spacer B-layers
in a consistent way with the magnetic structures of the A-layers. 
Similar oscillating inter-layer magnetic structures and
induced moments 
have been found in DMFT calculations,
which support the present FLEX study.~\cite{com:Peters}
Furthermore, such oscillations in the magnetic inter-layer coupling  
have also been commonly found in ferromagnetic superlattices.~\cite{pap:Grunberg1986,
pap:Parkin1990,pap:Parkin1991}
One intuitive understanding of this phenomena
is based on the RKKY interaction between magnetic 
layers separated by metallic spacer layers.~\cite{pap:Yafet1987,pap:Bruno1995}
The magnetic inter-layer coupling is asymptotically given by 
$\sim J\sin 2k_Fz/z^2$ with the Fermi wavenumber $k_F$ and coupling 
strength $J$.
In the present study, the system is close to half filling so that
the Fermi wavenumber of the $c$-electrons at $V=0$
along the $z$-axis is $\sim \pi/2$,
which leads to the above-mentioned periodicity in $\chi^f_s(q)$.

\begin{figure}
\begin{center}
\includegraphics[width=0.7\hsize,height=0.5\hsize]{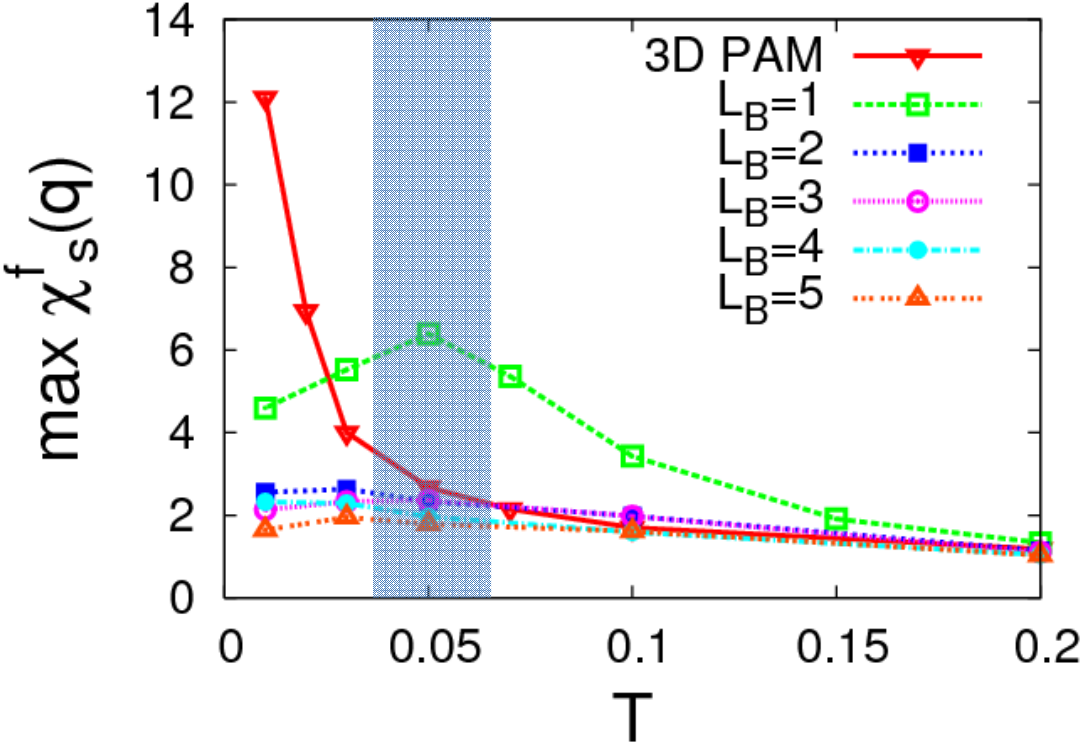}
\caption{
Temperature dependence of the maximum value of $\chi^f_s(i\omega_n=0,
\vecc{q})$ for $t^c_2=t^c_1$.
Shaded region indicates the characteristic temperature scale $T_0$.
}
\label{fig:maxchis}
\end{center}
\end{figure}

\begin{figure*}
\includegraphics[width=0.32\linewidth]{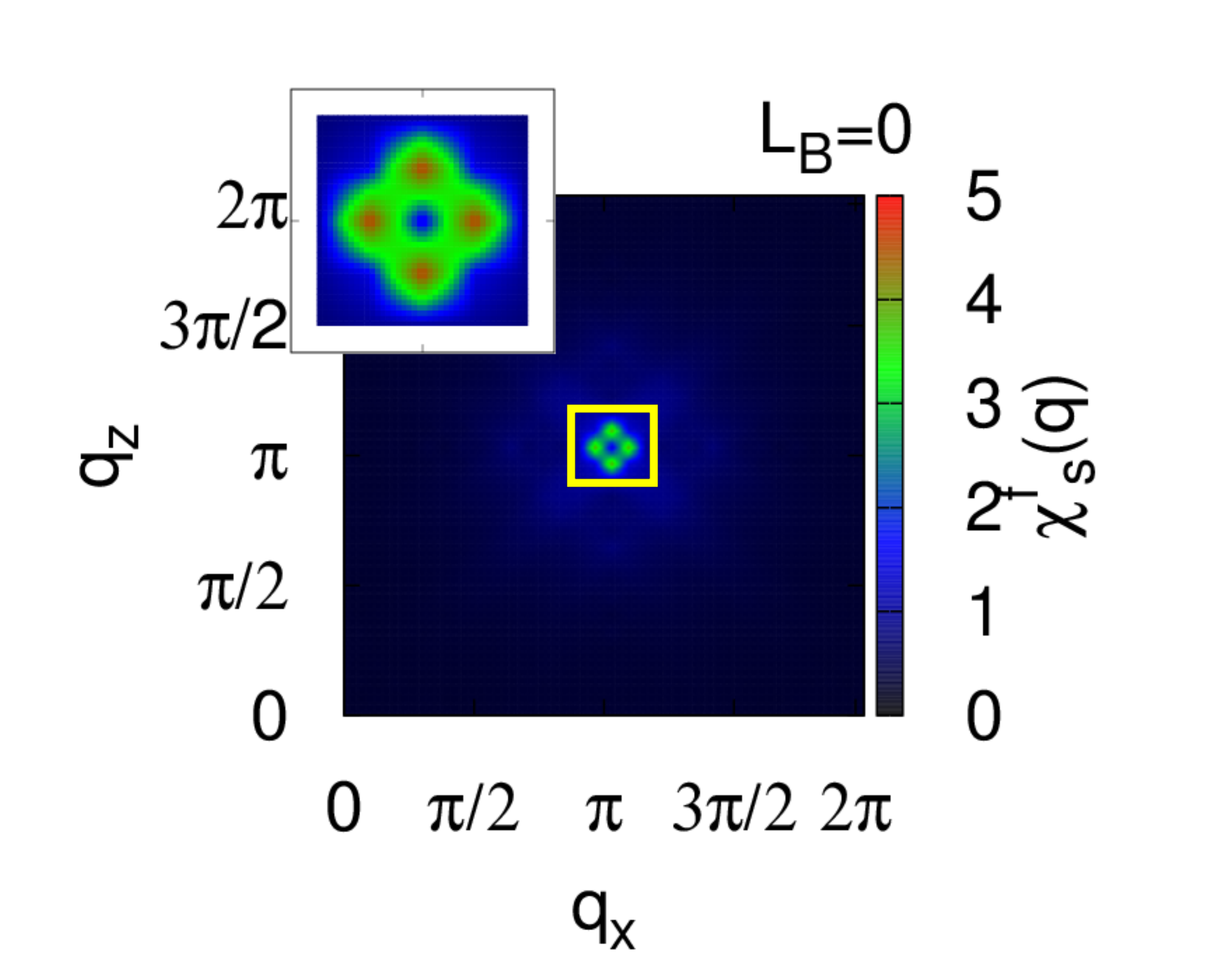}
\includegraphics[width=0.32\linewidth]{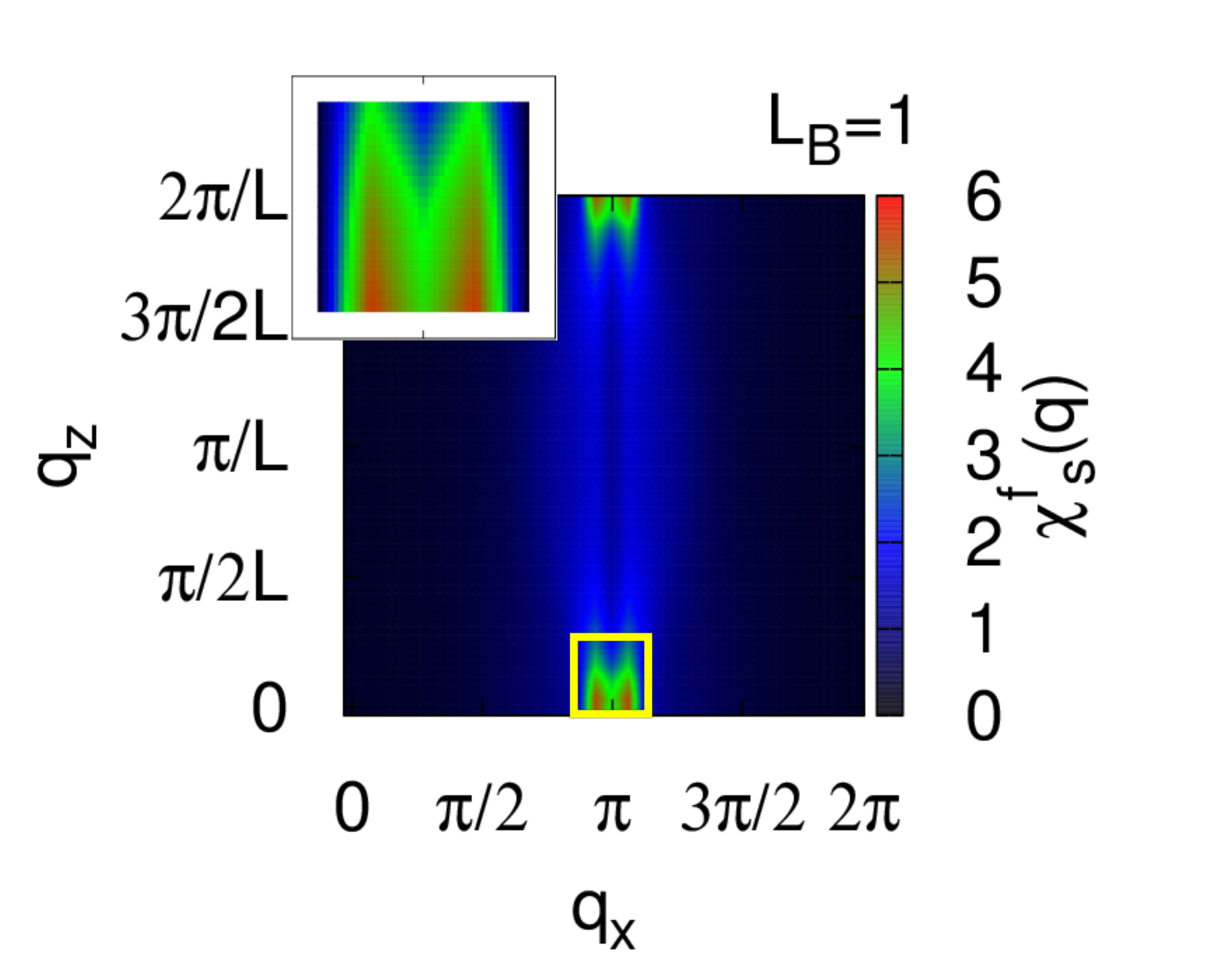}
\includegraphics[width=0.32\linewidth]{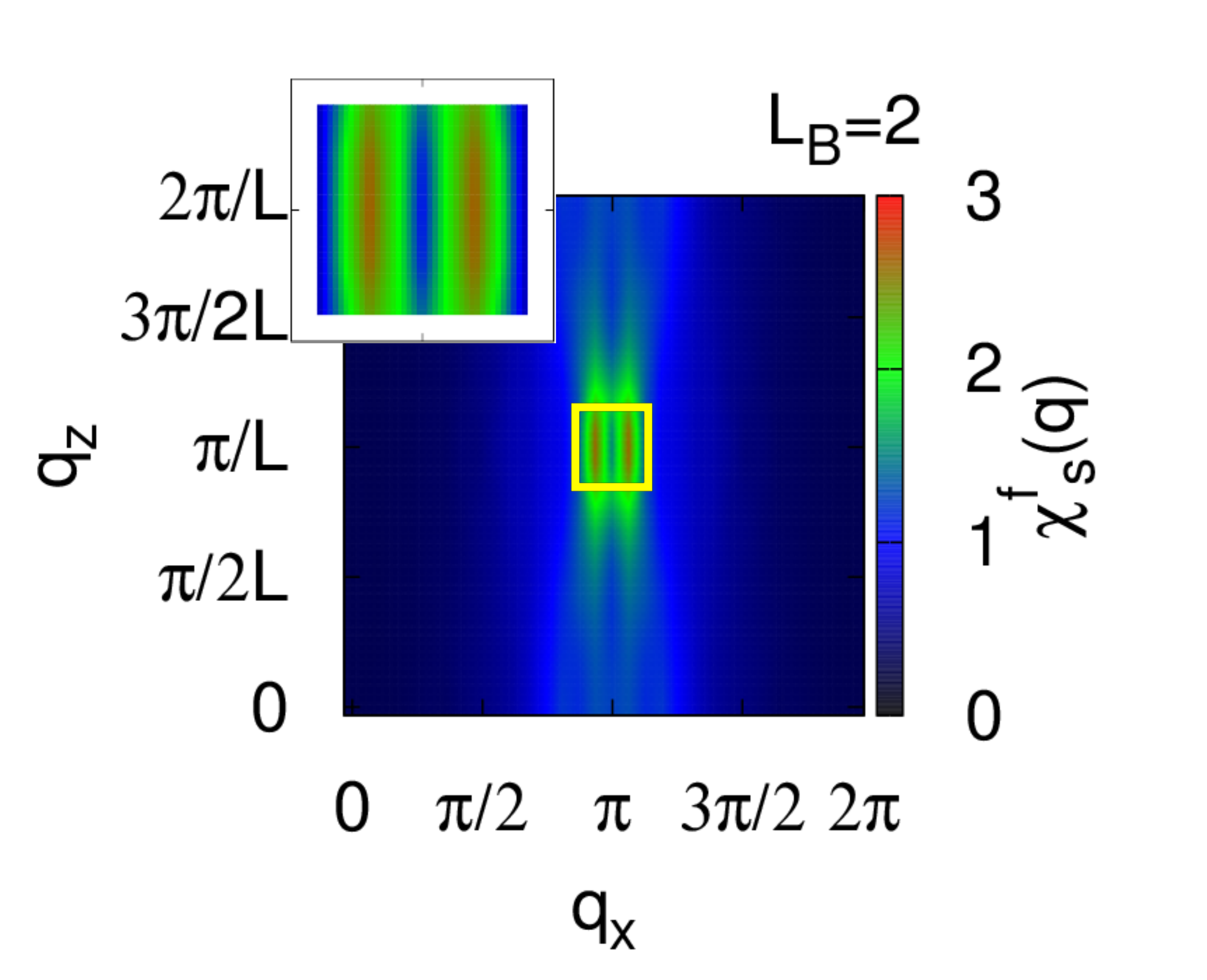}\\
\includegraphics[width=0.32\linewidth]{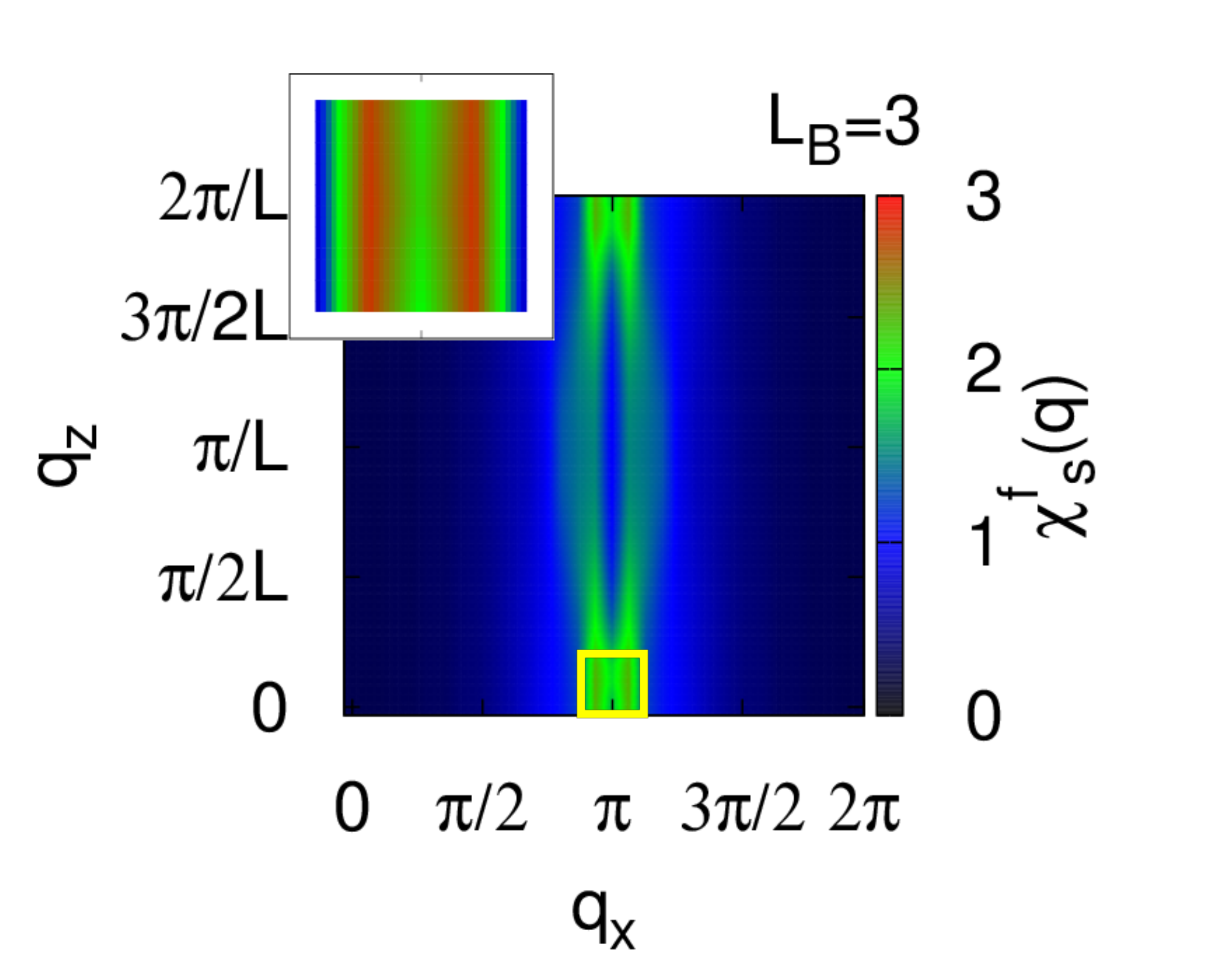}
\includegraphics[width=0.32\linewidth]{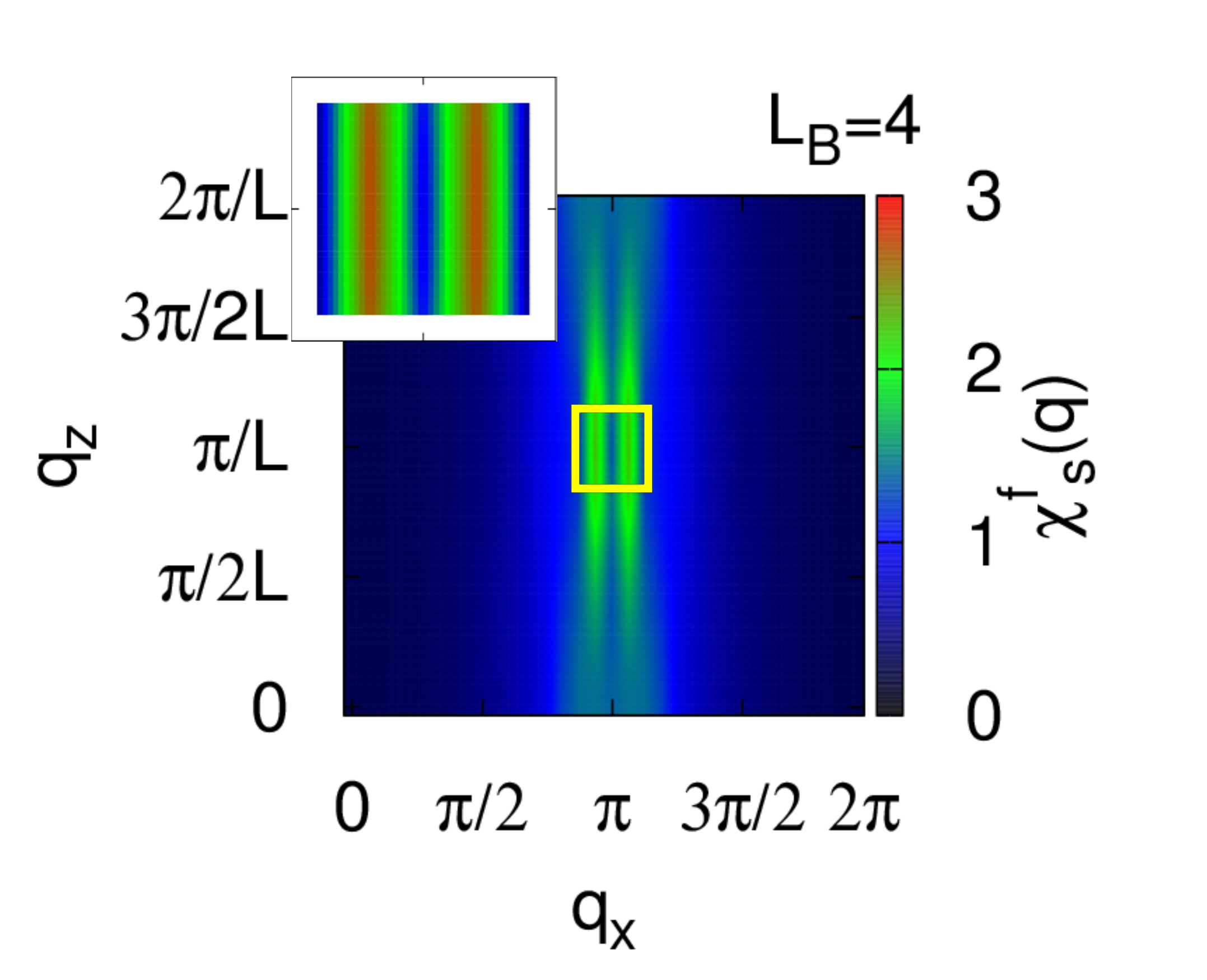}
\includegraphics[width=0.32\linewidth]{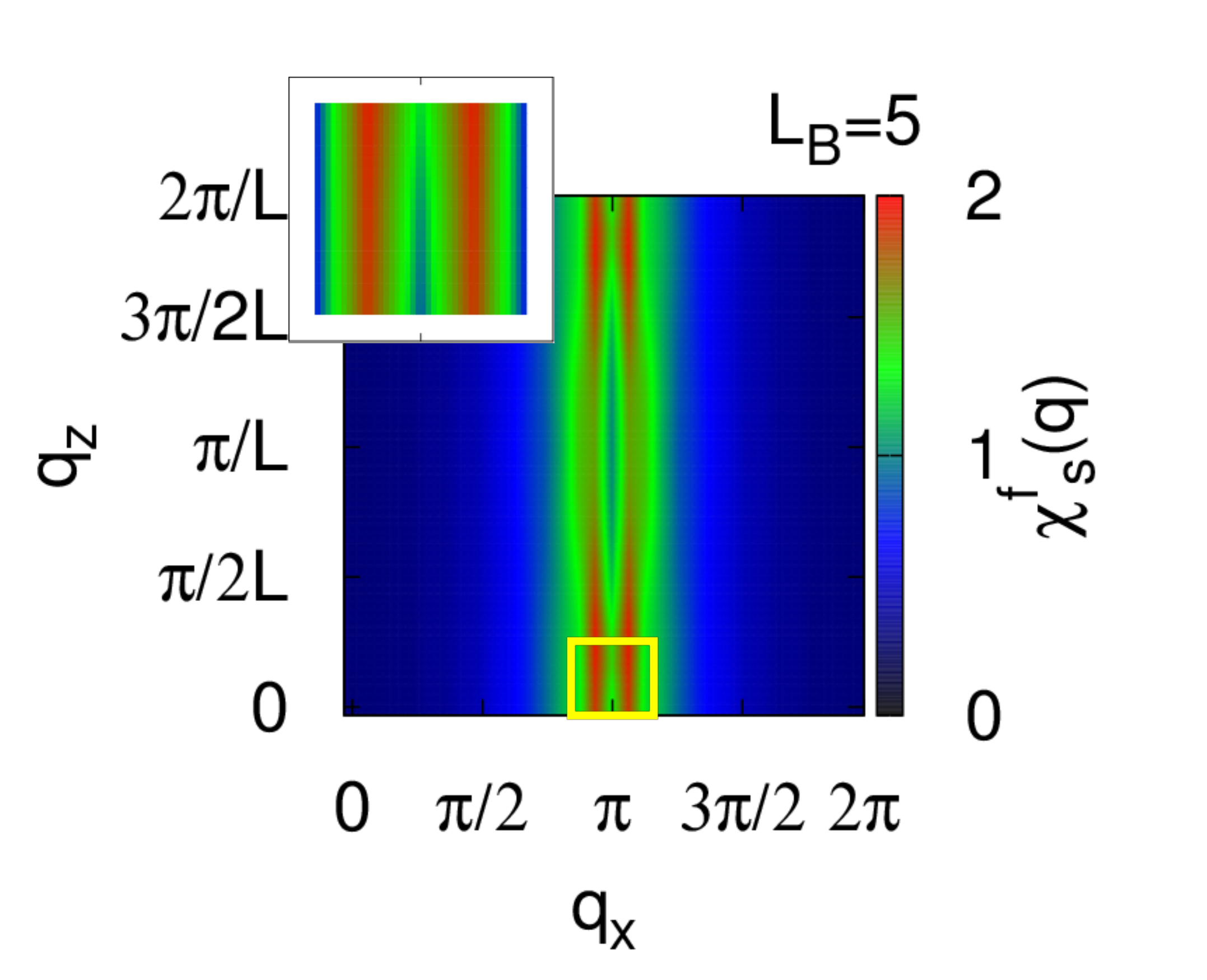}
\caption{
Spin susceptibility $\chi^f_s(i\omega_n=0,\vecc{q})$
at $T=0.03$
in the $xz$-plane at $q_y=\pi$ for $t^c_2=t^c_1$. The inset in each panel shows a magnification 
of the BZ as marked by the yellow square in the main plot.
}
\label{fig:chisT003}
\end{figure*}

\begin{figure}
\includegraphics[width=0.49\linewidth]{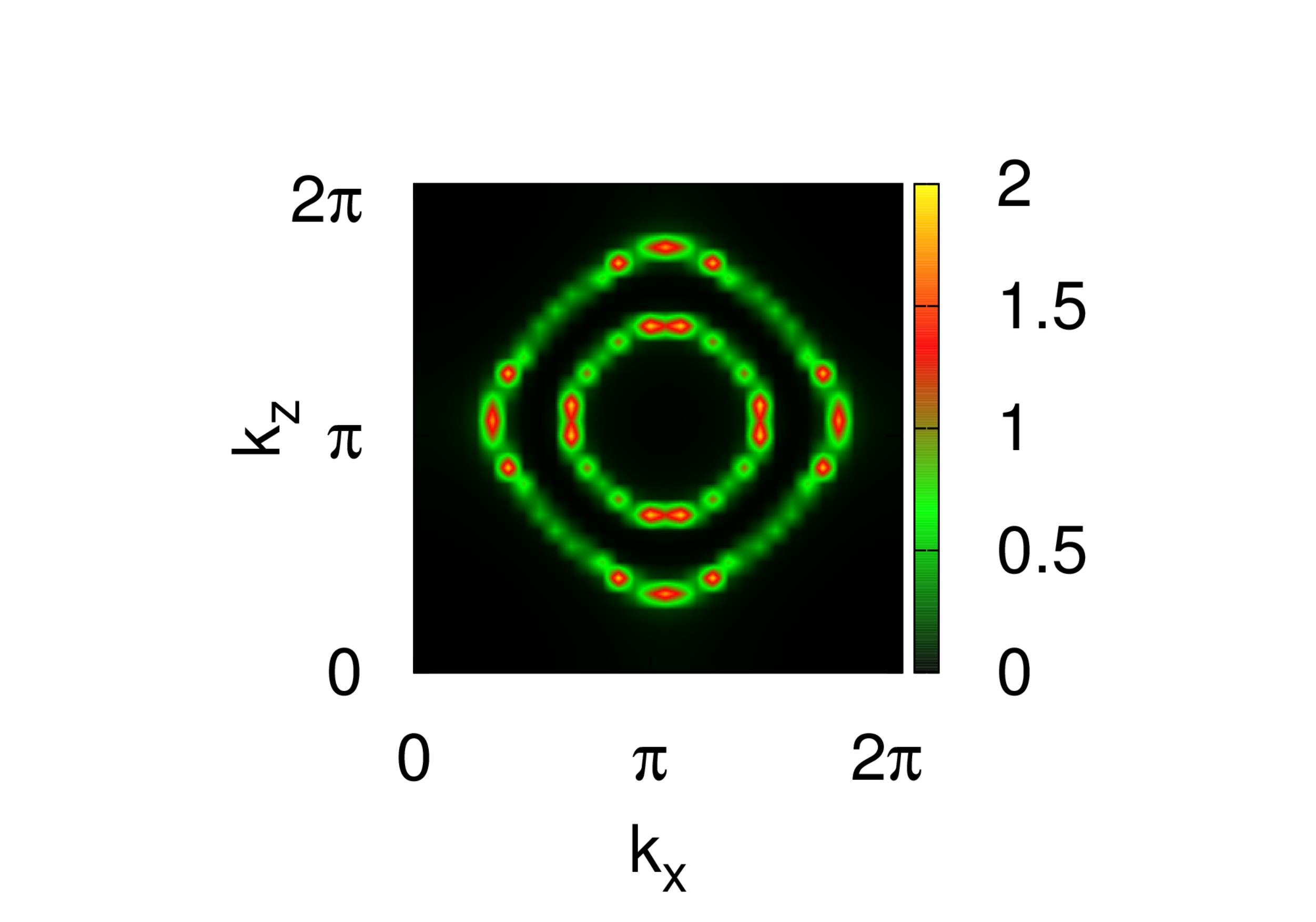}
\includegraphics[width=0.49\linewidth]{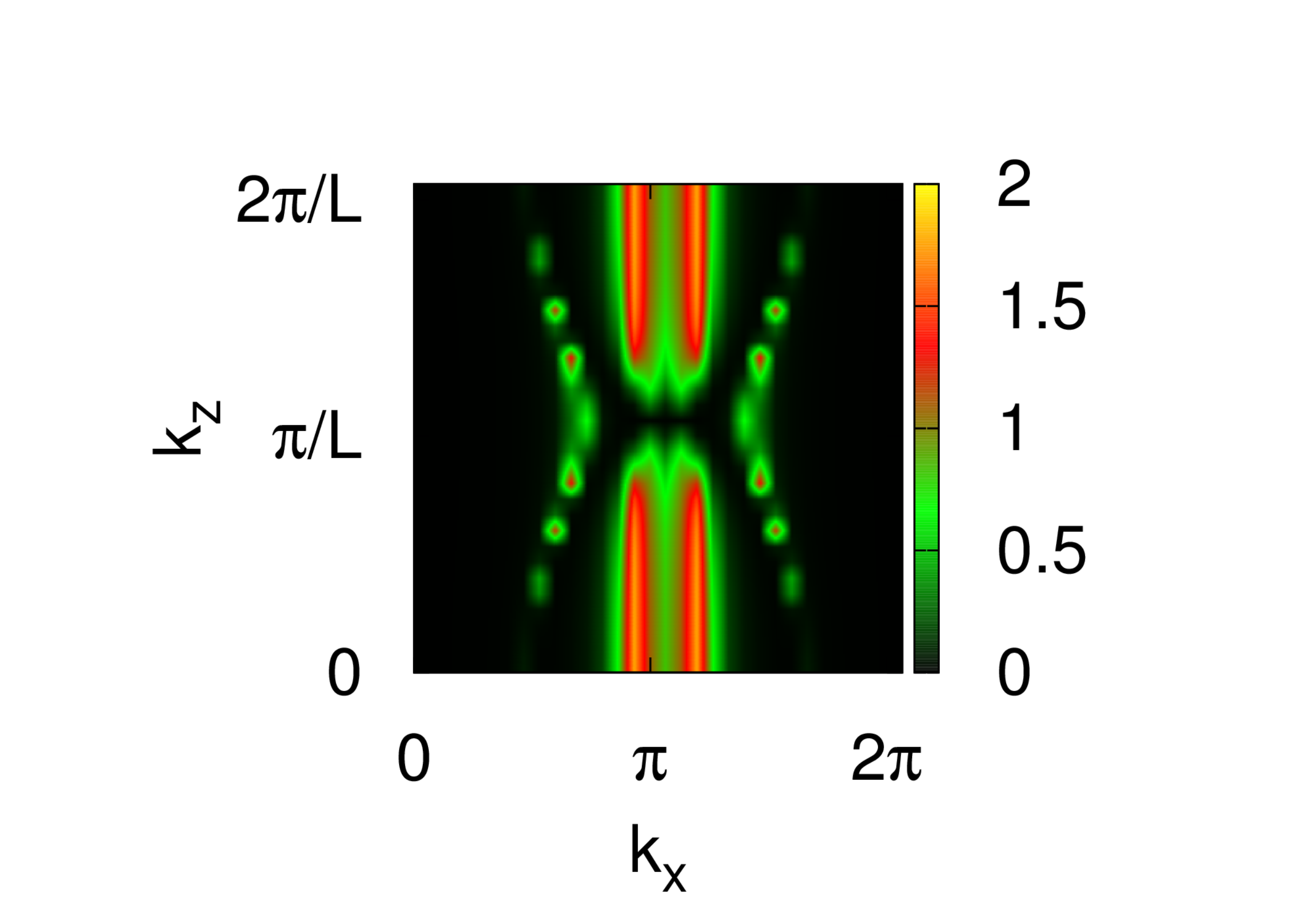}
\caption{Fermi surface as given by the
spectral function $(-1/\pi){\rm Im}G^f(i\pi T,\vecc{k})$
at a low temperature $T=0.03$ with $k_y=0$
for $L_B=0$ (left panel)
and $L_B=1$ (right panel).
}
\label{fig:FS}
\end{figure}

We can estimate the strength of the spin fluctuations by the 
maximum value $\chi^f_s(i\omega_n=0,\vecc{Q})$, 
which we show in Fig.~\ref{fig:maxchis}.
We want to note here, that the magnitude of this maximum value is strongly parameter dependent, 
because already small changes in the density of states can lead to a substantial enhancement
of the Stoner factor (Eq.~\ref{eq_sus}), if the system is close to magnetic
criticality.
A general behavior observed in our calculations is that the susceptibility $\chi^f_s$ for the 3D bulk system
 (without superlattice structure, $L_B=0$) strongly increases
below a characteristic temperature $T_0$, 
indicated by the shaded region in the figure, $T_0\sim 0.05$.
At the same temperature, the single peak in $\chi^f_s$, which is present at high temperature (Fig.~\ref{fig:chisT01}),
is split as shown in Fig.~\ref{fig:chisT003}.
Due to the hybridization between $c$- and $f$-electrons, the Fermi surface is split and the nesting properties are changed at low temperatures 
as seen in Fig.~\ref{fig:FS}.
Thus,
the spin fluctuations are strongly affected by $V$
for $T<T_0$ where heavy fermions with long lifetime 
are well formed within
the present FLEX calculations.

In the $f$-electron superlattice, $L_B\neq0$, 
the Fermi surface 
differs strongly from the 3D bulk system.~\cite{pap:Tada2013}
Particularly, a $q_z$-dependence of $\chi^f_s(q)$ arises only from $t^c_2$ through $V$,
because a direct hopping between different A-layers separated by
the spacer B-layers is forbidden in the present model.
A general behavior observed for the superlattice is that, similar to the bulk system, the single peak in the susceptibility at high temperature is split
into four peaks at low temperatures, $T<T_0\sim0.05$, as shown in Fig.~\ref{fig:chisT003}.
However, the abrupt increase in the susceptibility, which is observed in the bulk system at low temperature,
is cut off in the superlattice around $T_0$ and the susceptibility decreases for $T<T_0$, as shown in Fig.~\ref{fig:maxchis}.
While at high temperature, $T>T_0$, the Fermi surface is mainly determined by the $c$-electrons, 
below $T_0$ the $f$-electrons and thus the superlattice structure become important. 
As a consequence, 
for temperatures below $T_0$ the nesting of the Fermi surface in the superlattice is
changed 
and $\vecc{Q}\sim(\pi,\pi,\pi)$ is no longer a
good nesting vector as can be seen in Fig.~\ref{fig:FS}. 
Therefore, as $T$ is decreased and the Fermi surface becomes
affected by the superlattice structure through $V$,
the lack of good nesting properties of the Fermi surface 
cuts off the enhancement of the 
spin fluctuations. 
We note that the large value of max$[\chi_s^f]$ for $L_B=1$ originates
in a large density of states near the Fermi energy due to the superlattice
structure~\cite{pap:Peters2013} which leads to a substantial enhancement
in the Stoner factor, as mentioned above. However, while this strong enhancement at finite
 temperature for $L_B=1$ is strongly parameter dependent, the decrease of the susceptibility in the superlattice below $T_0$ has been observed 
for a wide range of the parameters.

\begin{figure}
\begin{center}
\includegraphics[width=0.7\hsize,height=0.5\hsize]{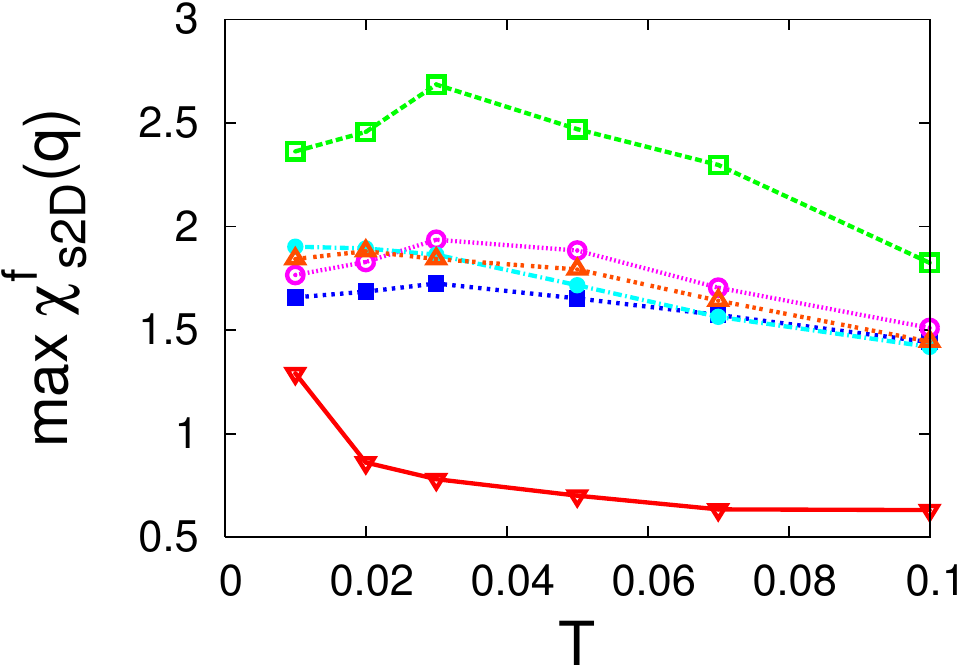}
\caption{
Temperature dependence of the maximum value of $\chi^f_{s{\rm 2D}}
(i\omega_n=0,
\vecc{q}_{\parallel})$ for $t^c_2=t^c_1$. Symbols are the same as
in Fig.~\ref{fig:maxchis}
}
\label{fig:chis2D}
\end{center}
\end{figure}

As shown
in Fig. \ref{fig:chisT003}, spin fluctuations in the superlattice are 
smeared out at low temperatures within the $q_z$ direction. 
The $q_z$-dependence of $\chi^f_s(q)$ becomes weaker as $L_B$ is increased. 
Thus, spin fluctuations become more 2-dimensional-like when $L_B$ is
increased.
However, this also means that the peak height of $\chi^f_s(q)$ in the
3D Brillouin zone is not an appropriate measure for the strength of the
spin fluctuations when $L_B$ is large.
In order to estimate the strength of the spin fluctuations,
we consider an effective 2D spin susceptibility
\begin{align}
\chi^f_{s{\rm 2D}}(i\omega_n,\vecc{q}_{\parallel})
=\frac{1}{N_z}\sum_{q_z}\chi^f_s(i\omega_n,\vecc{q}).
\end{align}
$\chi^f_{s{\rm 2D}}(q_{\parallel})$ has its maximum 
at $\vecc{Q}_{\parallel}\sim(\pi,\pi)$ for any $L_B$ in the present model.
We show the temperature dependence of the maximum values of
$\chi^f_{s{\rm 2D}}(q_{\parallel})$
in Fig.~\ref{fig:chis2D}.
Although max$[\chi^f_{s{\rm 2D}}]$ in the superlattice $(L_B\ge1)$
does not show a divergence at low temperature, 
this effective 2D spin susceptibility is clearly enhanced compared to
the 3D PAM ($L_B=0$) for the calculated temperature range.
We note that
the change in dimensionality of the spin fluctuations from 3D to 2D
has also been experimentally observed in the CeIn$_3$/LaIn$_3$ superlattice
as the CeIn$_3$-layer thickness ($L_A$ in our model) was tuned with a 
fixed LaIn$_3$-layer thickness ($L_B$).~\cite{pap:Shishido2010}
Although we cannot directly compare our results to the experiments,
the suppression of the magnetic order in the superlattice and the
calculated 2D-like character of the spin fluctuations are
consistent with the experiments.

\begin{figure*}
\includegraphics[width=0.32\linewidth]{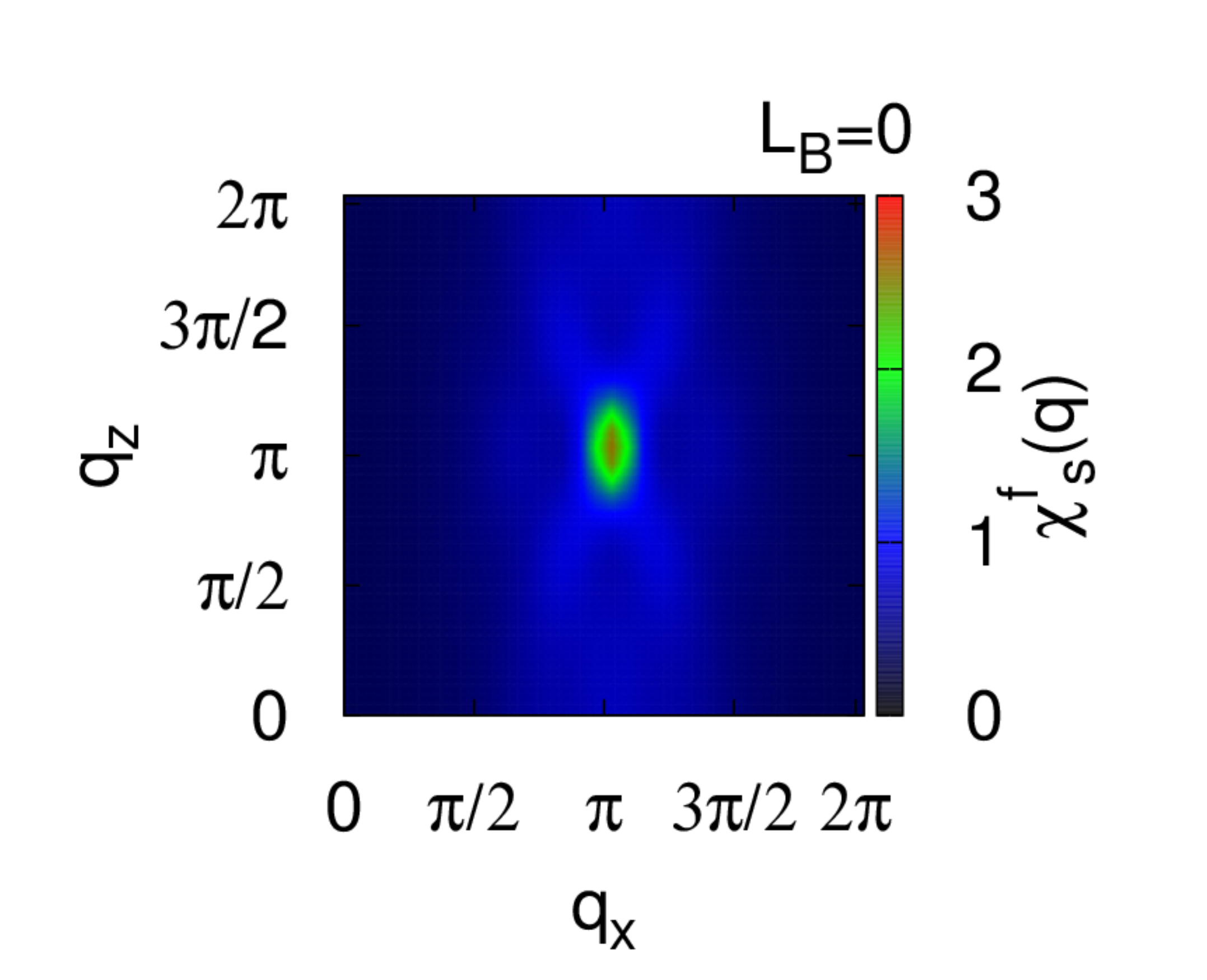}
\includegraphics[width=0.32\linewidth]{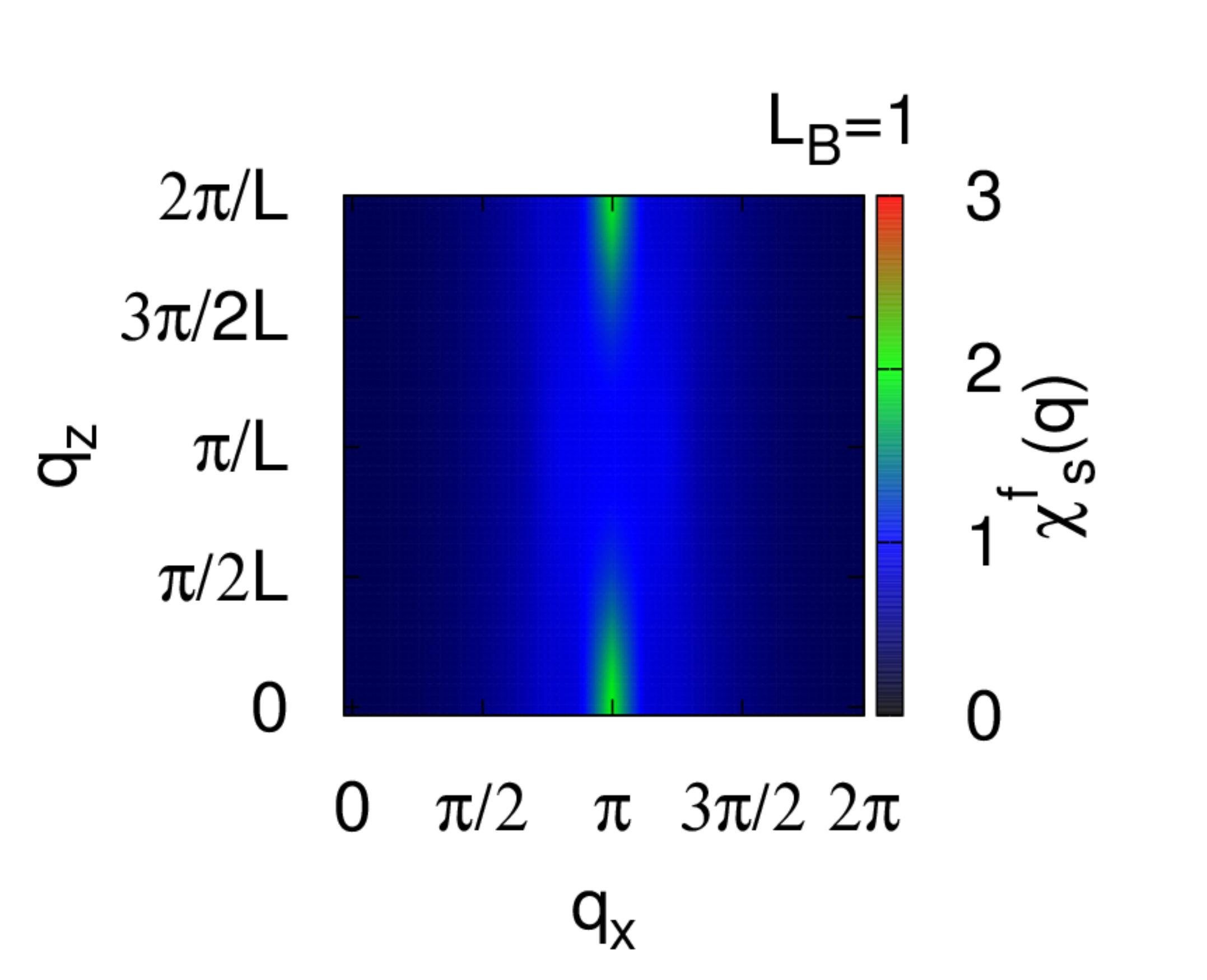}
\includegraphics[width=0.32\linewidth]{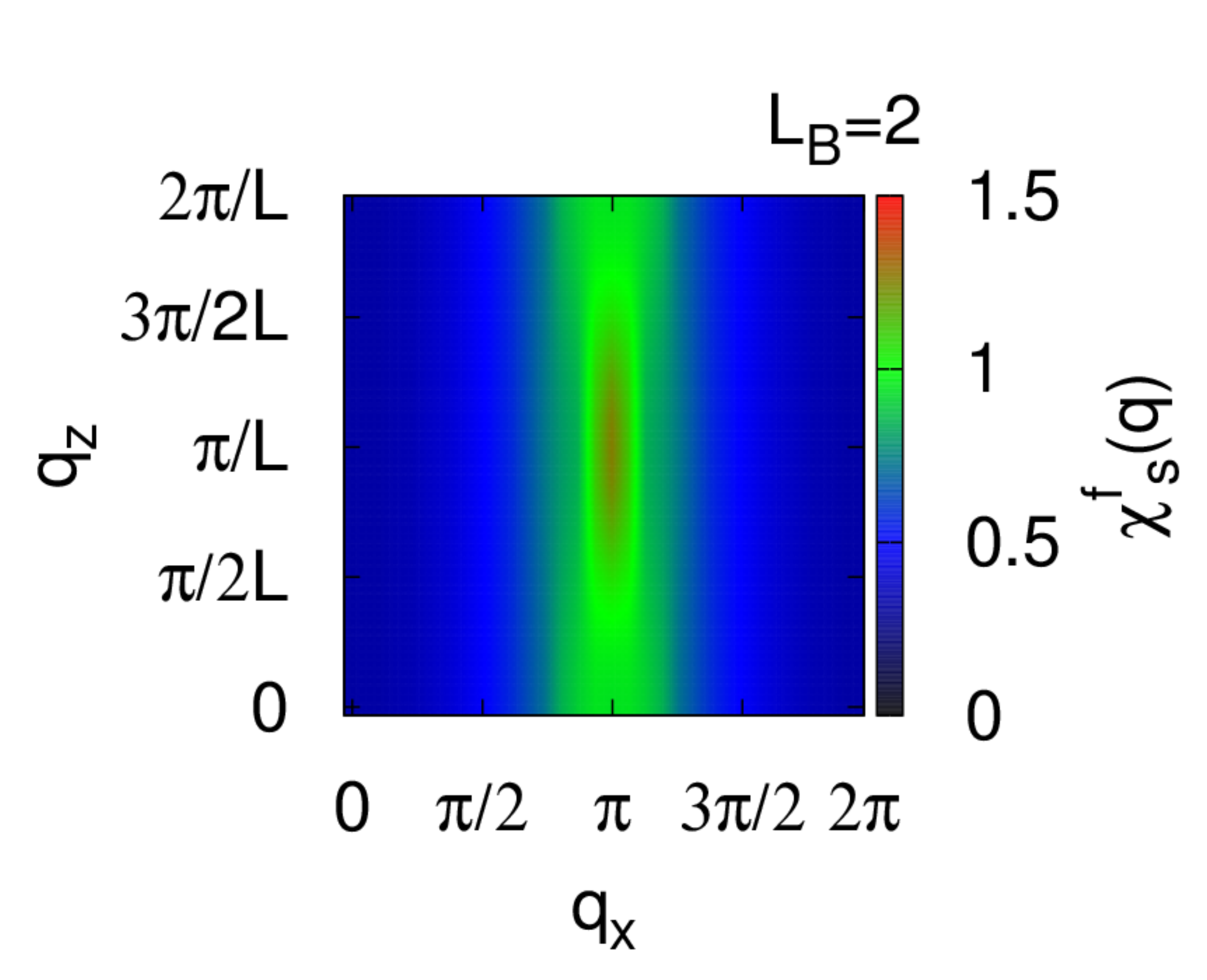}\\
\includegraphics[width=0.32\linewidth]{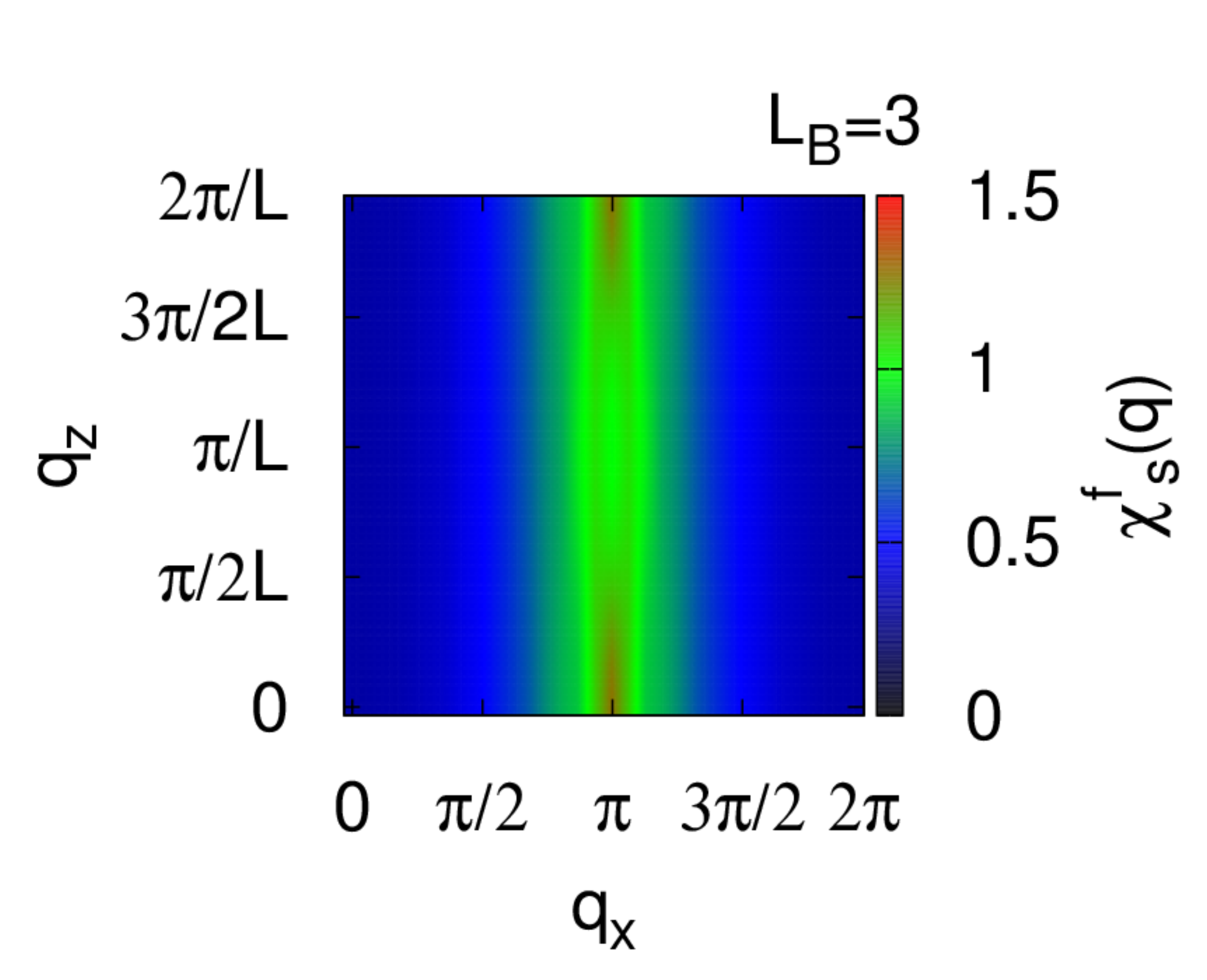}
\includegraphics[width=0.32\linewidth]{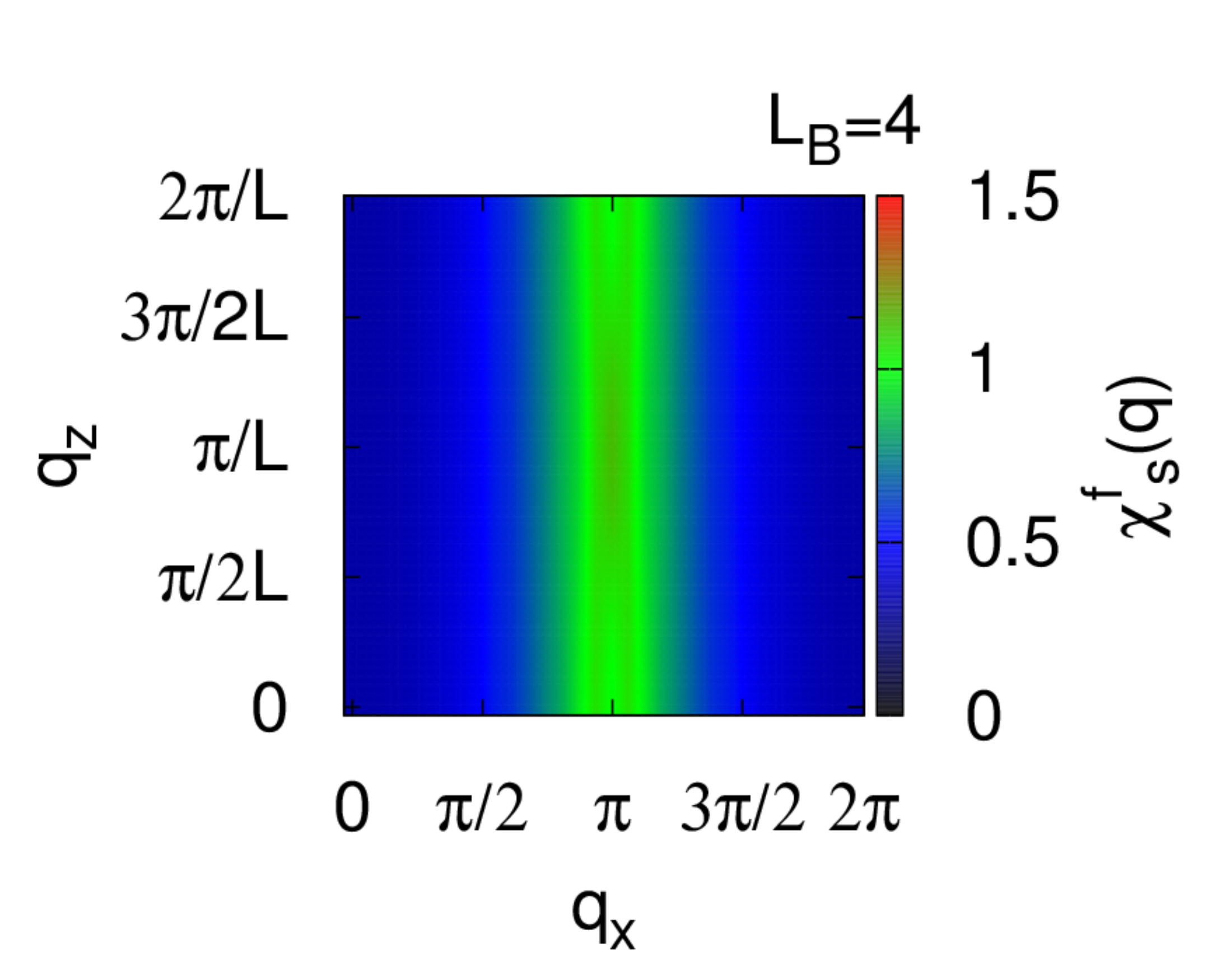}
\includegraphics[width=0.32\linewidth]{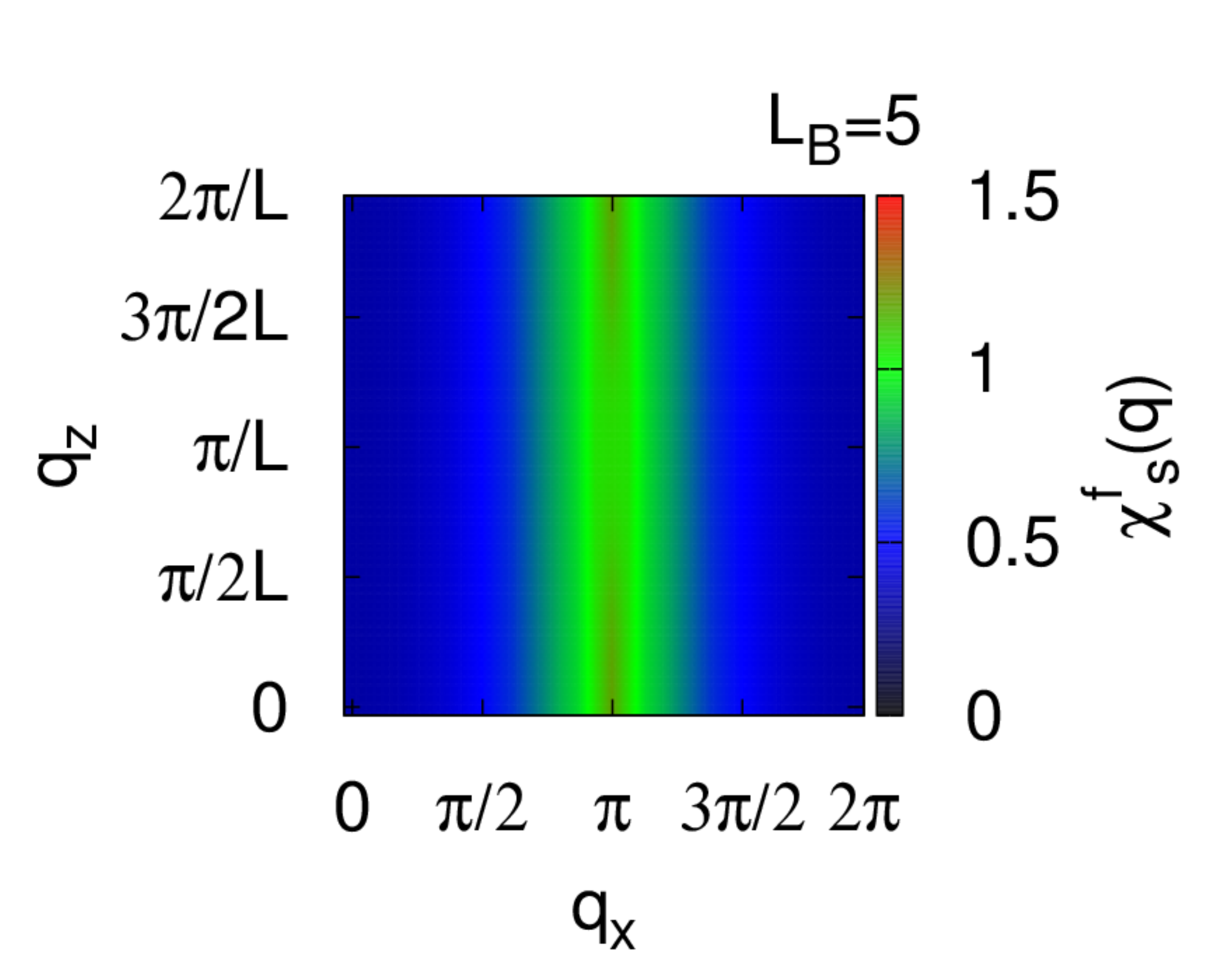}
\caption{
Spin susceptibility $\chi^f_s(i\omega_n=0,\vecc{q})$
at $T=0.1$
in the $xz$-plane at $q_y=\pi$ for $t^c_2=0.5t^c_1$.
}
\label{fig:chisT01_2}
\end{figure*}

\begin{figure}
\begin{center}
\includegraphics[width=0.7\hsize,height=0.5\hsize]{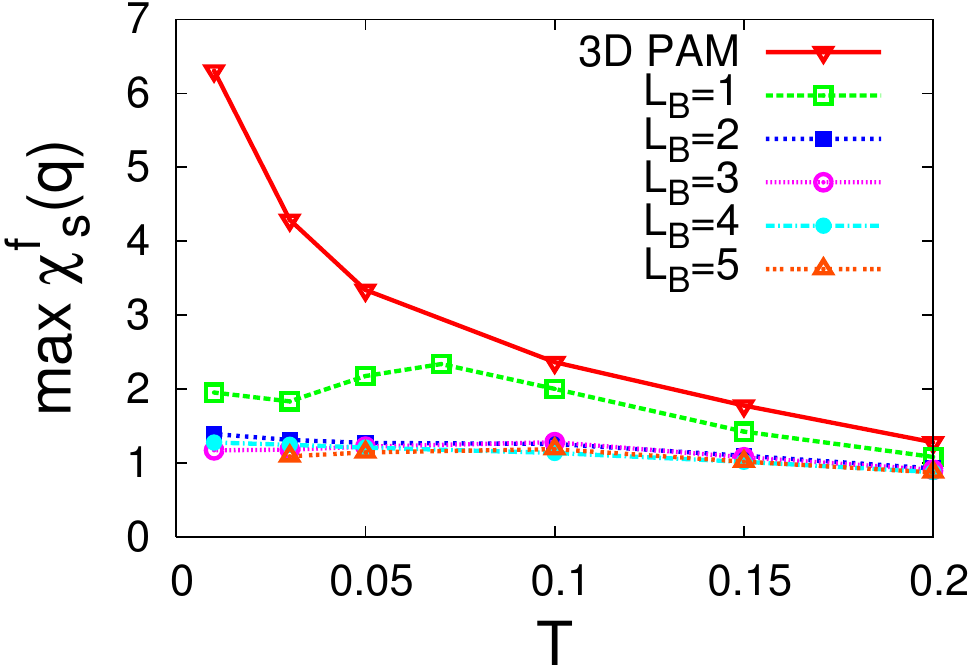}
\caption{
Temperature dependence of the maximum value of $\chi^f_s(i\omega_n=0,
\vecc{q})$ for $t^c_2=0.5t^c_1$.
}
\label{fig:maxchis_2}
\end{center}
\end{figure}

Up to now, we have analyzed the magnetic susceptibility for an isotropic  model. However, CeCoIn$_5$ 
exhibits a cylindrical Fermi surface~\cite{pap:Shishido2002}.
In order to investigate effects of an anisotropy in the
original 3D model,
we consider a system with 
a more 2D-like set of hopping parameters, $t^c_{2}=0.5t^c_1$.
Similarly to the isotropic hopping parameter set, 
we observe even-odd oscillations of the peak positions when $L_B$ is changed,
as shown in Fig.~\ref{fig:chisT01_2}.
We show the
temperature dependence of the maximum values of $\chi^f_s(T)$ within the 3D Brillouin zone
in Fig.~\ref{fig:maxchis_2}.
The susceptibility of the 3D PAM ($L_B=0$), max$[\chi^f_s(T)]$, behaves again monotonically 
and rapidly grows at low temperature. Furthermore, any increase in  max$[\chi^f_s(T)]$ for
 the superlattice, $L_B>0$, is again cut off at low temperatures. 
However, because of the anisotropic parameter set, the bulk system, $L_B=0$, already includes strong 2D spin fluctuations, 
and the effective 2D spin susceptibility $\chi^f_{s{\rm 2D}}$ (not shown in the paper)
is less enhanced in the superlattices.
Comparing the results for the two parameter sets, $t^c_2=t^c_1$
and $t^c_2=0.5t^c_1$, 
we find that the impact of the $f$-electron confinement to the A-layers 
is stronger when the hopping parameters are more 3-dimensional.
This is reasonable,
because in the limit of decoupled layers, $t^c_2\rightarrow0$,
the superlattice structure does not play a role at all.
 We have confirmed this tendency by performing similar calculations for different 
hopping parameters.

\section{superconductivity}
\label{sec:superconductivity}
\begin{figure}[htbp]
\begin{center}
\includegraphics[width=0.7\hsize,height=0.5\hsize]{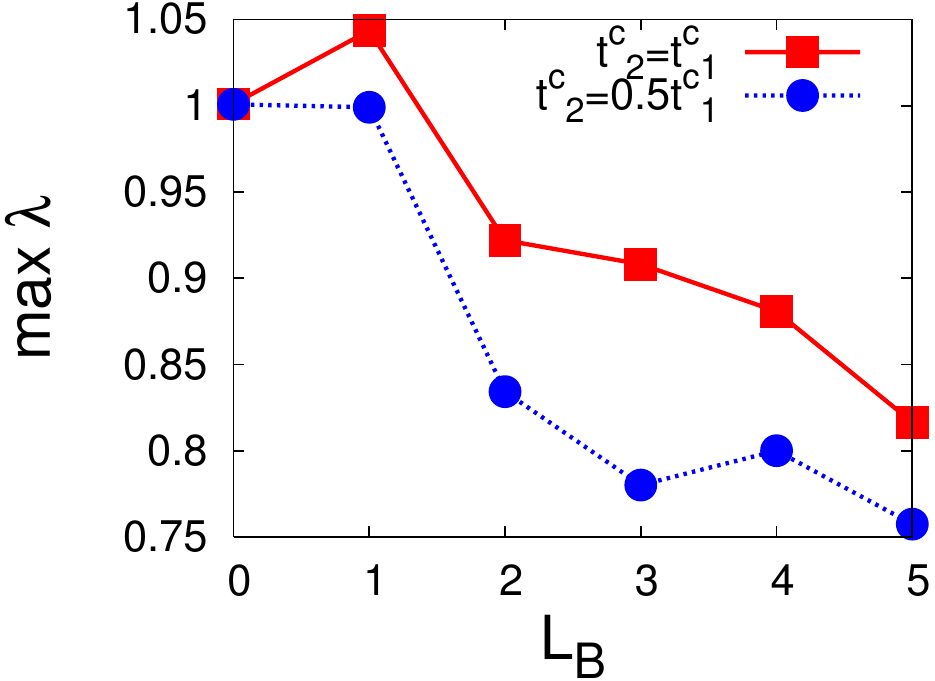}
\caption{
Maximum eigenvalue $\lambda$ for the $s$-wave superconductivity
at fixed $T=0.02$.
}
\label{fig:lambda_s}
\end{center}
\end{figure}

Finally, we investigate the impact of the increased 2D spin fluctuations in the superlattice on the superconductivity
by solving the Eliashberg equation (\ref{eq:Eliashberg}).
First, in order to understand the proximity effect in our model,
we consider $s$-wave superconductivity by replacing
$V^f_s(q)$ in Eq. (\ref{eq:Eliashberg}) by 
\begin{align}
V^f_{\rm s-wave}(q)=-U_0\theta(\omega_D-|\omega_n|),
\end{align}
where we fix $\omega_D=1.0$ and $U_0$ is tuned so that
$\lambda=1$ when $L_B=0$ and $T=0.02$. We then solve the Eliashberg equation without selfenergy.
Thus, electron correlations are not included in this calculation.
In Fig.~\ref{fig:lambda_s}, 
we show the maximum eigenvalues, max$[\lambda]$,
of the 
Eliashberg equation which corresponds to the strength of the superconducting instability. 
The largest eigenvalue, $\lambda$, rapidly decreases as $L_B$ is increased,
showing slight oscillations 
due to changes in the density of states at the Fermi energy
which depend on the details of the model parameters.
Thus, 
if conventional $s$-wave superconductivity, mediated by phonons, were realized 
in  CeCoIn$_5$/YbCoIn$_5$
superlattices, 
one can expect that $T_c$ is
strongly decreased in the superlattice. 
Such a suppression of the superconductivity 
has been commonly observed in superlattices composed of $s$-wave superconductors and normal metals.~\cite{book:Shinjo2002,book:ChangGiessen1985,pap:Jin1989,
pap:Barnerjee1984,
pap:Kanoda1986}
Experimentally, $T_c$ is lower in
all the previous conventional superlattices compared to the
 corresponding bulk systems.
It has theoretically been shown that the superconducting transition temperature
$T_c$ is suppressed in an exponential way in layered systems when the thickness of the
normal metal layer $L_N$
is increased,~\cite{pap:deGennes1963,pap:deGennes1964,pap:Werthamer1963,
pap:Werthamer1964}
\begin{align}
T_c(L_N)\simeq T_c(0)-\delta T_c\tanh (L_N/\xi_0),
\label{eq:proximity}
\end{align}
where $\delta T_c\simeq (T_c(L_N)/T_c(\infty))(T_c(0)-T_c(\infty))$ 
and $\xi_0$ is effective coherence length. 
This suppression of $T_c$ comes from the fact that the
pairing interaction which exists only in the superconductor layer mediates
superconductivity not only in the superconductor layer, but also 
in the normal metal layer.
Since this is a general property of the proximity effect,
one could naively expect a suppression of superconductivity also for 
$d$-wave states, which are mediated by the spin-fluctuations.

\begin{figure}
\begin{center}
\includegraphics[width=0.7\hsize,height=0.5\hsize]{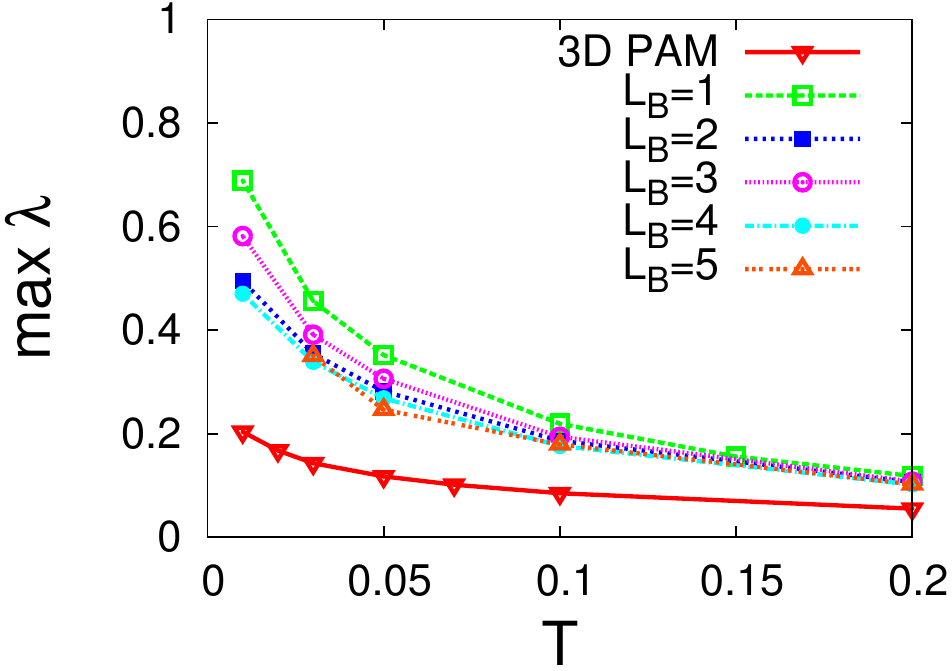}
\caption{
The maximum eigenvalue of the Eliashberg equation for the $d_{x^2-y^2}$-wave
superconductivity when $t^c_2=t^c_1$.
}
\label{fig:lambda}
\end{center}
\end{figure}

\begin{figure}[tp]
\includegraphics[width=0.49\linewidth]{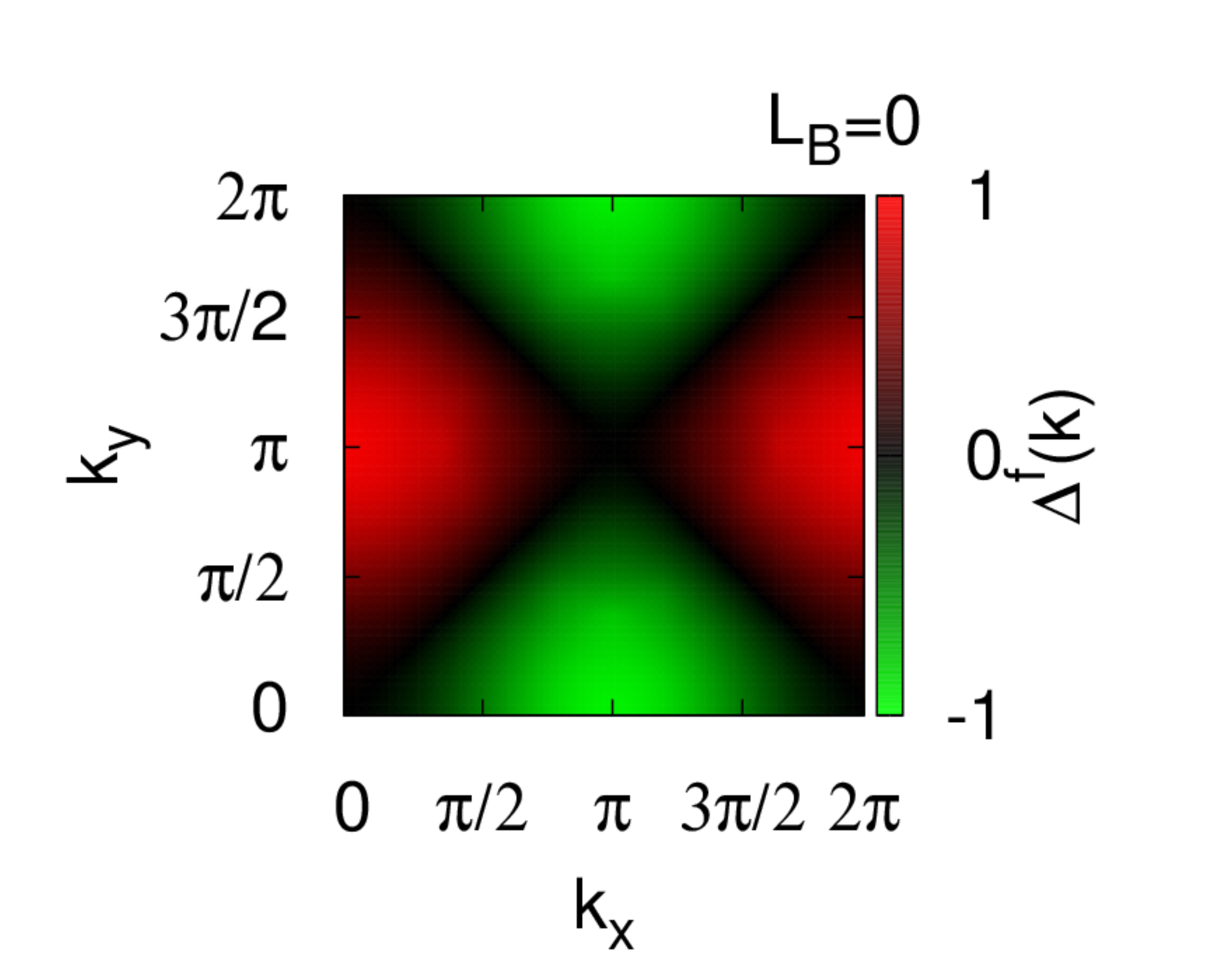}
\includegraphics[width=0.49\linewidth]{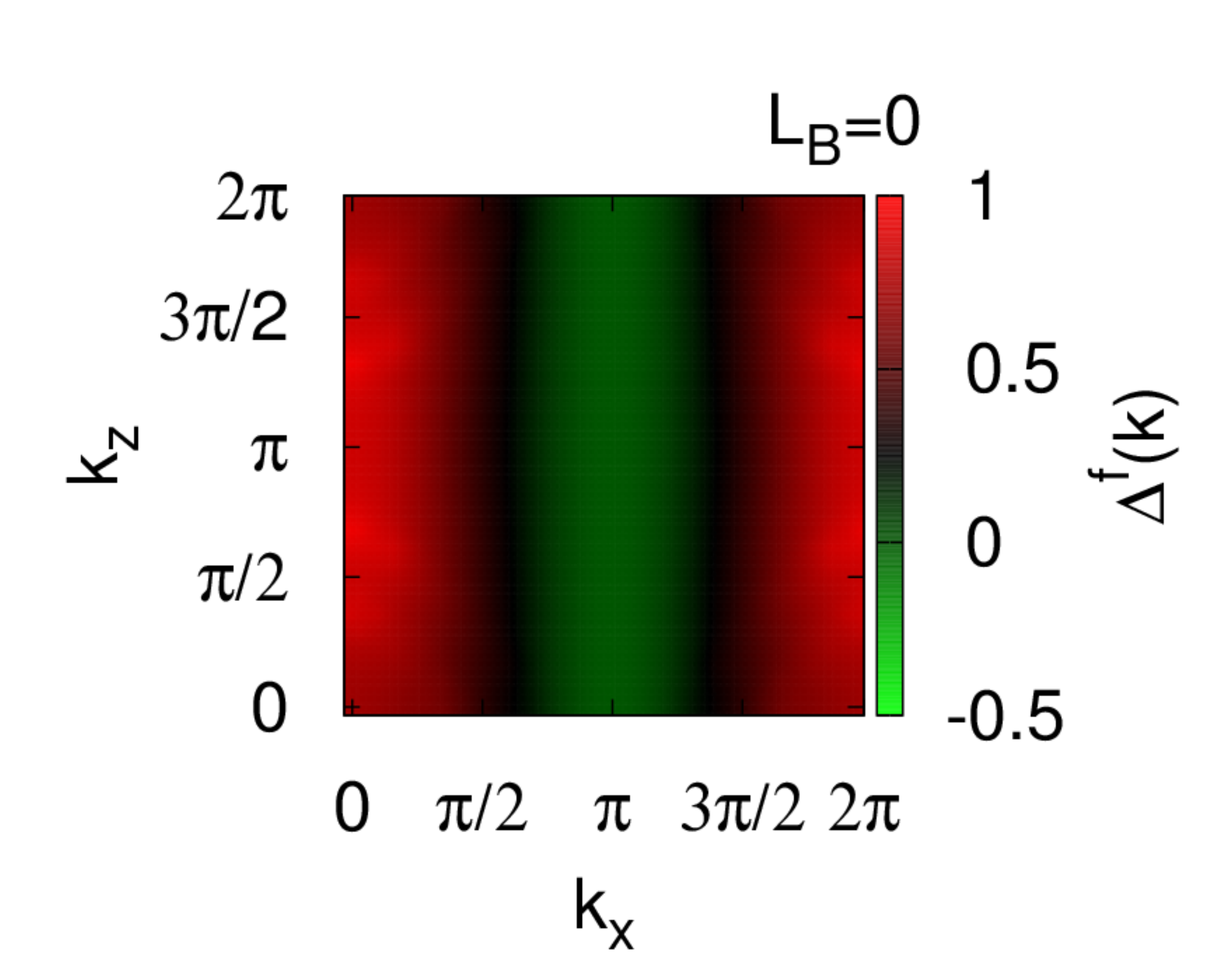}\\
\includegraphics[width=0.49\linewidth]{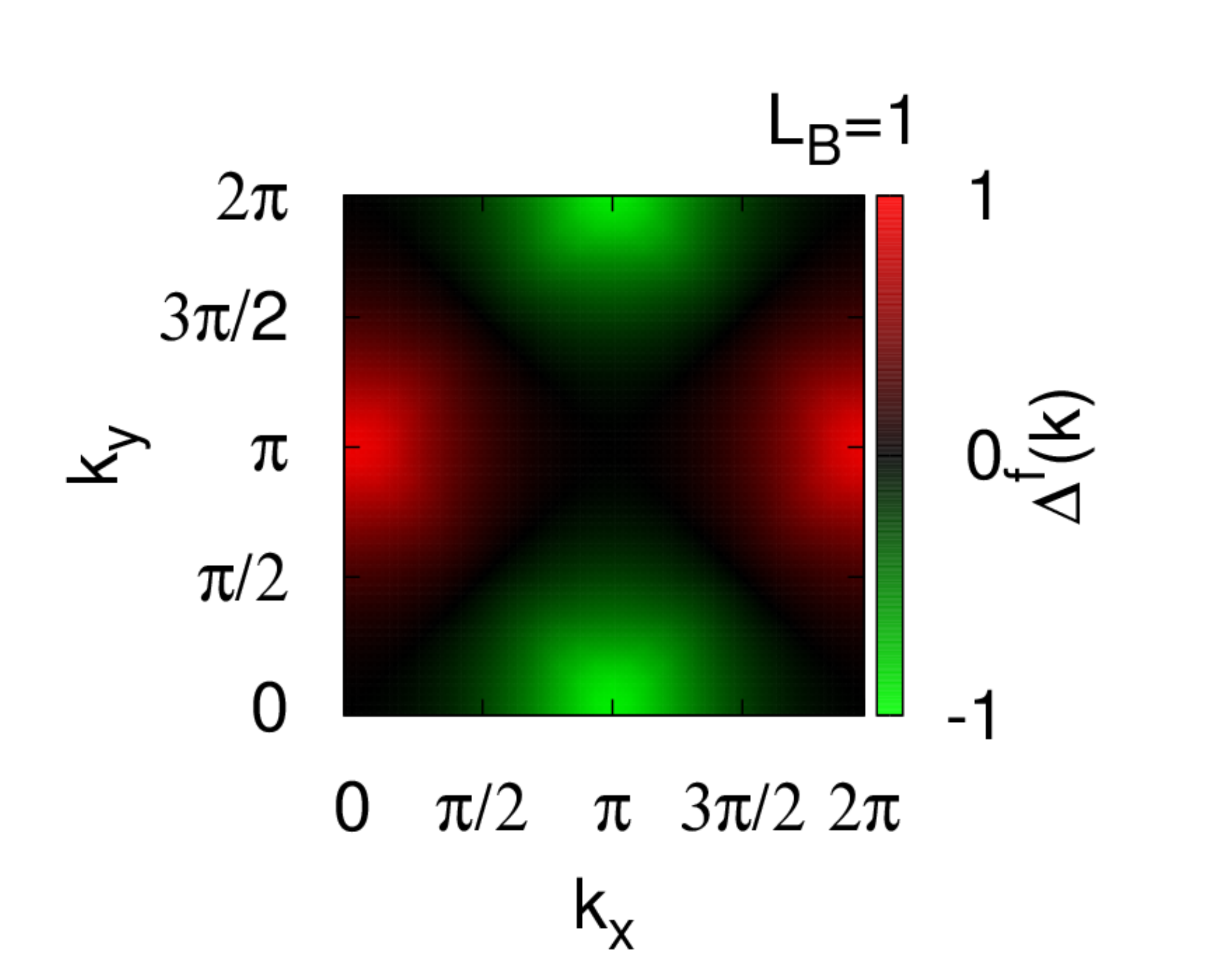}
\includegraphics[width=0.49\linewidth]{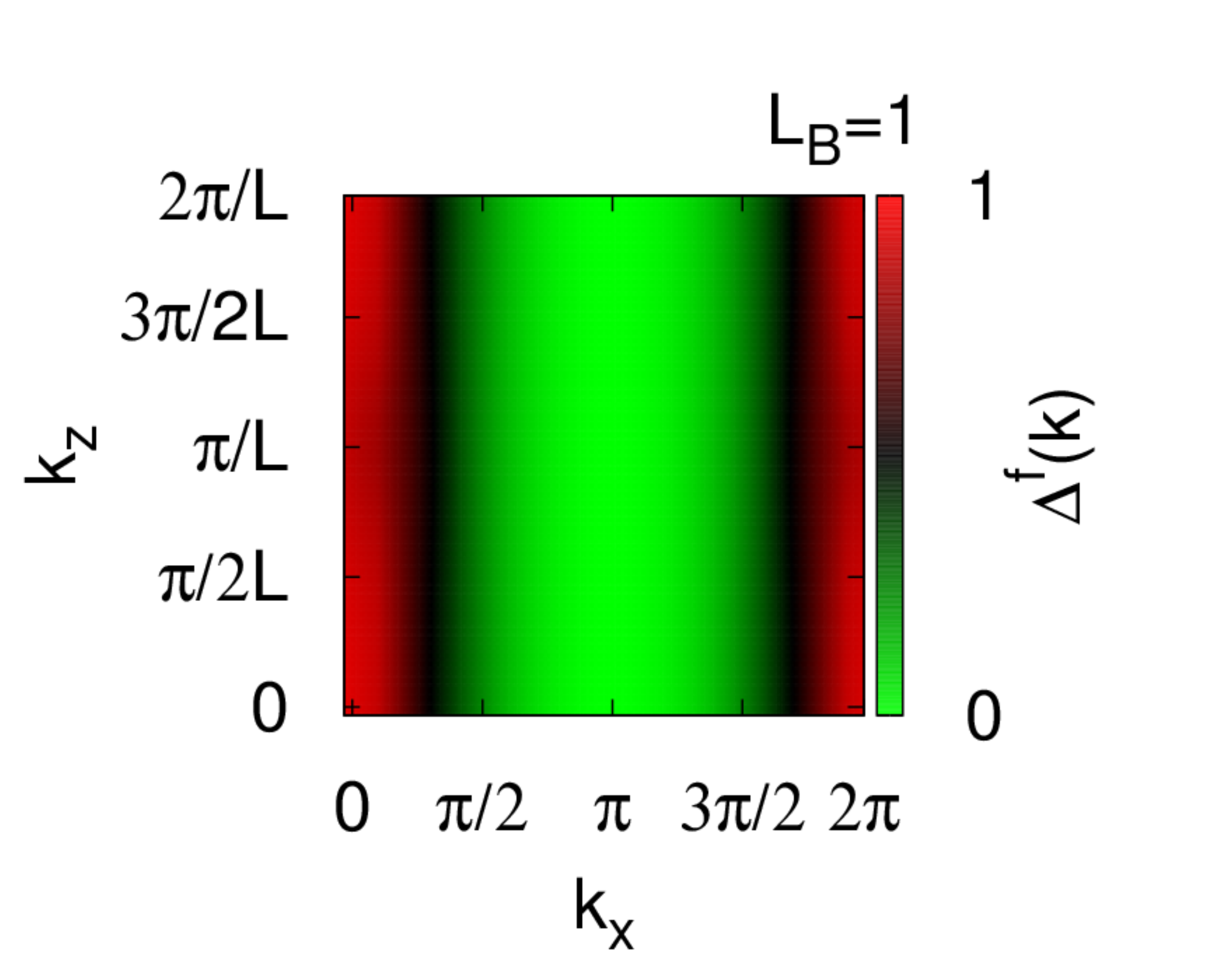}
\caption{
Gap functions $\Delta(k_x,k_y,k_z=0)$ and 
$\Delta(k_x,k_y=0,k_z)$ at $\omega_n=\pi T$
for $L_B=0,1$ when $t^c_2=t^c_1$ and $T=0.02$
in arbitrary units.
}
\label{fig:gap}
\end{figure}

In order to examine this further, we solve the Eliashberg equation with
the pairing interaction $V_s^f(q)$ corresponding to spin fluctuations
and the normal selfenergy
for $t^c_2=t^c_1$. 
Contrary to the $s$-wave superconductivity, as seen in Fig.~\ref{fig:lambda}, 
the maximum eigenvalues for the $d_{x^2-y^2}$-wave
superconductivity mediated by the spin fluctuations are {\it enhanced} 
in the $f$-electron superlattices. 
Although the eigenvalues, max$[\lambda]$, do not reach unity for the calculated temperature range,
these results suggest that $T_c$ for $d_{x^2-y^2}$-wave
superconductivity
 can be higher in the superlattice than in the bulk system.
This strong $d$-wave superconducting instability in the $f$-electron
superlattice can be understood
by focusing on the effective 2D spin fluctuations 
discussed in the
previous section.
In order to stabilize the $d_{x^2-y^2}$-wave superconductivity with 
$\Delta(k)\propto (\cos k_x-\cos k_y)$, 
the $q_z$-dependence in $V^f_s(q)$ is irrelevant
and we only need to consider the $q_xq_y$-dependence.
As exemplified in Fig.~\ref{fig:gap}, 
typical profiles of the $d_{x^2-y^2}$-wave gap functions are indeed
 $\Delta(k)\sim (\cos k_x-\cos k_y)$ and their 
$k_z$-dependence is weak for any $L_B$.
If we neglect the $k_z$-dependence in $|G^f(k)|^2$,
the Eliashberg equation is reduced to
\begin{align}
&\Delta^f(k_{\parallel})\simeq -\frac{T}{N_{\parallel}}
\sum_{k'_{\parallel}}V^f_{s{\rm 2D}}(k_{\parallel}-k_{\parallel}')
|G^f(k_{\parallel}')|^2\Delta^f(k_{\parallel}'),\\
&V^f_{s{\rm 2D}}(q_{\parallel})=\frac{1}{N_z}\sum_{q_z}V^f_s(q).
\end{align}
From this equation, it is clear that the most important part of the
pairing interaction is determined by $\chi^f_{s{\rm 2D}}(q_{\parallel})$
which is enhanced in the superlattice (Fig.~\ref{fig:chis2D}).
This enhancement of the effective pairing
interaction 
can lead to an increased $T_c$, which is a consequence of
the interplay between strong 
interaction among the $f$-electrons and the
confinement of them within the superlattice structure.
This is characteristic for the $f$-electron superlattice.
We note that a similar enhancement of superconductivity has been theoretically
 found in 3D bulk models when tuning the hopping parameter in the $z$-direction~\cite{pap:Monthoux2001,pap:Arita2000}.
  However, in these studies the proximity effect does not play a role. In our
   present study, superconductivity is enhanced as a result of a subtle
    interplay between the proximity effect and an increase of spin fluctuations
     in the superlattice.

\begin{figure}
\begin{center}
\includegraphics[width=0.7\hsize,height=0.5\hsize]{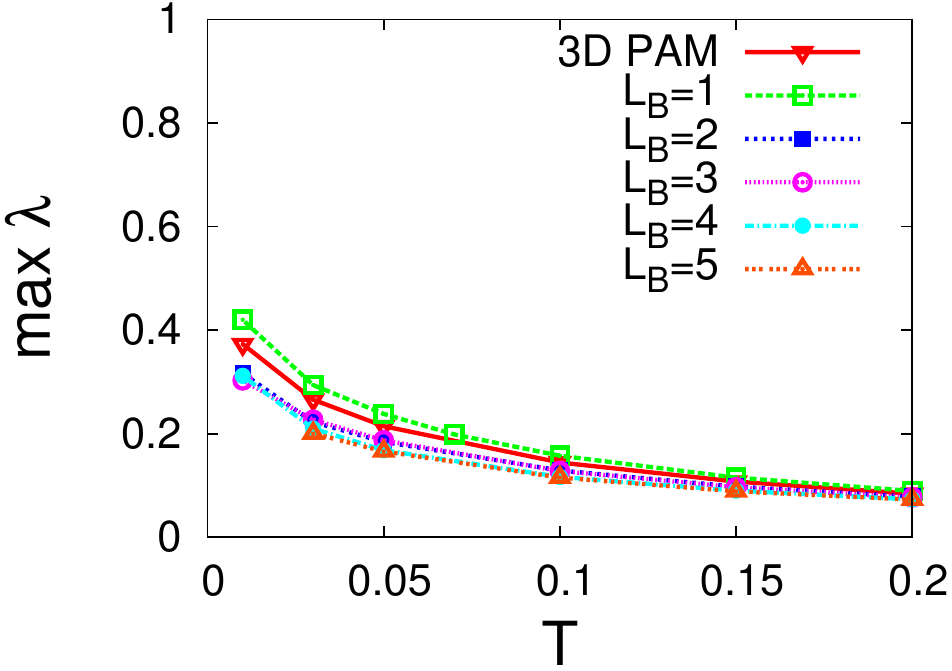}
\caption{
The maximum eigenvalue of the Eliashberg equation for the $d_{x^2-y^2}$-wave
superconductivity when $t^c_2=0.5t^c_1$.
}
\label{fig:lambda2}
\end{center}
\end{figure}

Results for max$[\lambda]$ when $t^c_2=0.5t^c_1$ are shown 
in Fig.~\ref{fig:lambda2}.
For this parameter set the enhancement of $\chi^f_{s{\rm 2D}}$ in the superlattice is weak. 
Therefore, also the maximum eigenvalue of the Eliashberg equation 
are only slightly increased in the superlattice.
$T_c$ in the superlattice would be similar to the bulk value
when the hopping parameters are 2D-like.
However, the enhancement of $\chi^f_{s{\rm 2D}}$
in the superlattice due to the $f$-electron confinement
is still important for these parameters. It almost cancels
the suppression of $d$-wave superconductivity due to the
proximity effect.
Therefore,
$T_c$ in the superlattice can remain as high as in the bulk $(L_B=0)$
even for large $L_B$.

Experimentally it has been observed that $T_c$ is lower in the CeCoIn$_5$/YbCoIn$_5$ superlattice
 than in bulk  CeCoIn$_5$ with $T_c\simeq 2.3$(K).
We think that this can be explained by two reasons:
First, the Fermi surface of
CeCoIn$_5$ is cylindrical~\cite{pap:Shishido2002} and thus would be better described 
by the anisotropic parameter set in our calculations.
Second,
disorder effects seem to be strong for thin CeCoIn$_5$-layer 
superlattices.~\cite{pap:Mizukami2012}
In the experiments, the thickness of the YbCoIn$_5$-layers has been fixed and
the number of the CeCoIn$_5$-layers has been tuned.
We expect that if the thickness of the YbCoIn$_5$-layer is changed
with a fixed CeCoIn$_5$-layer width, 
the behavior of $T_c$ will deviates from conventional normal-metal/superconductor superlattices.

\section{summary}
\label{sec:summary}
We have investigated the $f$-electron superlattice based on
FLEX. We found that the nature of the spin fluctuations is modified by the
superlattice structure and the $\vecc{Q}$-vectors corresponding to the maximum in the susceptibility
depend on the width of the spacer layers, similar to
 conventional magnetic superlattices.
While the strength of the 3D spin fluctuations, characterized by 
max$[\chi^f_s(q)]$ in the 3D Brillouin zone, is suppressed in the superlattice
because good nesting properties of the Fermi surface are lost,
effective 2D fluctuations are enhanced because of reduced 
dimensionality.
These enhanced spin fluctuations can lead to higher $T_c$ in the case of
$d_{x^2-y^2}$-wave superconductivity in the superlattice
than in the bulk compounds, which is in sharp contrast to all the
conventional superlattice superconductors.
We hope that these results will lead to further experiments analyzing $T_c$ 
in clean $f$-electron superlattices.

\section*{ACKNOWLEDGEMENT}
We thank
Y. Matsuda, T. Shibauchi, H. Ikeda, 
S. Fujimoto, N. Kawakami, and Y. Yanase for valuable discussions. 
This work is supported by JSPS/MEXT KAKENHI
Grant Number 26800177 (Y.~T.).
RP thanks RIKEN for support through its FPR Program.

\bibliography{sulattice}

\end{document}